\documentclass{aa}  
\usepackage{graphicx}
\usepackage{txfonts}
\usepackage{xcolor}
\usepackage{lipsum}
\usepackage{subcaption}         
\usepackage{lscape}      
\usepackage{placeins} 
\usepackage[rightcaption]{sidecap}

\usepackage[normalem]{ulem}

\begin{document}

\title{Orbital motion and dynamical mass of the complex periodic variable binary system 2MASS\,J05082729$-$2101444}
   
\titlerunning{Orbital motion and dynamical mass of a young binary system}
\authorrunning{Curiel et al.}

\author{
S.~Curiel\inst{1}\thanks{\email{scuriel@astro.unam.mx}}, 
G.\,N.~Ortiz-Le\'on\inst{2}, 
V.\,J.\,S.~B\'ejar\inst{3,4}, 
D.~Vigan\`o\inst{5,6}, 
J.\,M.~Girart\inst{5,6}, 
S.~Kaur\inst{5,6}, 
Y.~Shan\inst{7,8}, 
F.~Murgas\inst{3,4}, 
M.~Zechmeister\inst{8},
P.\,J.~Amado\inst{9}, 
J.\,A.~Caballero\inst{10}, 
Th.~Henning\inst{11}, 
E.~Ilin\inst{12}, 
D.~Montes\inst{13}, 
J.\,C.~Morales\inst{5,6}, 
\`O.~Morata\inst{5,6}, 
M.~P\'erez-Torres\inst{9,14,15}, 
A.~Quirrenbach\inst{16}, 
A.~Reiners\inst{8}, 
I.~Ribas\inst{5,6}, 
\'A.~S\'anchez-Monge\inst{5,6}, 
A.~Schweitzer\inst{17}, 
J.\,I.~Vico~Linares\inst{18}, 
M.\,R.~Zapatero Osorio\inst{10}
}

\institute{Instituto de Astronom{\'\i}a, Universidad Nacional Aut\'onoma de M\'exico (UNAM), Apartado postal 70-264, Ciudad de M\'exico, M\'exico
\and
Instituto Nacional de Astrof{\'\i}sica, \'Optica y Electr\'onica, Apartado postal 51 y 216, 72000 Puebla, M\'exico
\and
Instituto de Astrof\'isica de Canarias, 38205 La Laguna, Tenerife, Spain
\and
Departamento de Astrof\'isica, Universidad de La Laguna, 38206 La Laguna, Tenerife, Spain
\and
Institut de Ci\`encies de l'Espai (ICE-CSIC), Campus UAB, Carrer de Can Magrans s/n, 08193 Cerdanyola del Vall\`es, Barcelona, Spain
\and
Institut d'Estudis Espacials de Catalunya (IEEC), Esteve Terradas 1, edifici RDIT, Par Mediterrani de la Tecnologia, Campus Baix Llobregat - UPC, 08860 Castelldefels, Barcelona, Spain
\and
The Science Library, Universitet i Oslo, Moltke Moes vei 35, 0851 Oslo, Norway
\and
Institut f\"ur Astrophysik und Geophysik, Georg-August-Universit\"at G\"ottingen, Friedrich-Hund-Platz 1, 37077 G\"ottingen, Germany
\and
Instituto de Astrof\'isica de Andaluc\'ia, CSIC, Glorieta de la Astronom\'ia s/n, 18008 Granada, Spain
\and
Centro de Astrobiolog\'ia, CSIC-INTA, Camino Bajo del Castillo s/n, Campus ESAC, 28692 Villanueva de la Ca\~nada, Madrid, Spain
\and
Max-Planck-Institut f\"ur Astronomie, K\"onigstuhl 17, 69117 Heidelberg, Germany
\and
Netherlands Institute for Radio Astronomy, ASTRON, Dwingeloo, Netherlands
\and
Departamento de F\'isica de la Tierra y Astrof\'isica \& IPARCOS (Instituto de F\'isica de Part\'iculas y del Cosmos), Facultad de Ciencias F\'isicas, Universidad Complutense de Madrid, Plaza de Ciencias 1, 28400 Madrid, Spain
\and
Departamento de F\'isica Te\'orica, Universidad de Zaragoza, 50009 Zaragoza, Spain
\and
School of Sciences, European University Cyprus, Diogenes street, Engomi, 1516 Nicosia, Cyprus
\and
Landessternwarte, Zentrum f\"ur Astronomie der Universit\"at Heidelberg, K\"onigstuhl 12, 69117 Heidelberg, Germany
\and
Hamburger Sternwarte, Gojenbergsweg 112, 21029 Hamburg, Germany
\and
Centro Astron\'omico Hispano en Andaluc\'ia, Observatorio Astron\'omico de Calar Alto, Sierra de los Filabres, 04550 G\'ergal, Almer\'ia, Spain
}

\date{Received 16 March 2026 / Accepted 01 May 2026}

\abstract
{}
{This study focuses on the low-mass binary 2MASS J05082729$-$2101444 (2M0508--21), one of the few known radio-loud members of the complex periodic variable sample. Our aim is to use very long baseline interferometry to constrain the orbit.}
{We observed the system with the Very Long Baseline Array (VLBA) in three epochs at a frequency of 4.85\,GHz, which provides an angular resolution of about 3\,mas. We combined the three radio astrometric observations, 119 RVs (60 VIS and 59 NIR) obtained with the CARMENES high-resolution spectrograph over a period of 8.1 years, and a relative astrometric measurement of an archival $H$-band Keck NIRC adaptive optics image to fit the orbital motion of the binary system.}
{The VLBA observations resolved the binary system and show emission from both stellar components, with similar flux-density levels (0.34--0.67\,mJy), and showing slight temporal flux variations. The emission appears quiescent, with no significant circular polarization, and with no flare events. We obtained an orbital motion fit of the binary system, which shows an eccentric orbit ($e$ = 0.71), an orbital period of 2.19\,yr, and a semimajor axis of 26.964 mas (1.3\,au).}
{The VLBA observations made it possible to resolve the binary system and identify both stars as radio-loud sources. The combined fit shows that 2M0508--21 is an M-dwarf binary with a total dynamical mass of $0.459\pm0.007\,$M$_{\odot}$, assuming \textit{Gaia}'s parallax. This mass is slightly higher than those estimated from the luminosity and theoretical evolutionary models. The upper limit of the circular polarization at 4.85\,GHz ($\lesssim$10\%), the persistence of the quiescent emission, and the relatively low brightness temperatures are consistent with a gyro-synchrotron or synchrotron origin of the radio emission. Further VLBA observations are needed to obtain the individual masses of the stars, as well as to verify {\em Gaia}'s parallax of the system. A complete characterization of the system will help improve evolutionary models for young objects at the substellar boundary.}


\maketitle
\nolinenumbers

\section{Introduction}\label{sec:intro}

The number of known M-dwarf stars emitting nonthermal radiation at 0.1--10\,GHz frequencies is steadily increasing. Although these radio-loud stars still represent a minority within their class, probably due to their intrinsically faint and variable radio emission, the increasing sensitivity of radio interferometers and the coverage of their wide-field surveys or systematic campaigns  
can now enable us to reveal the features of the radio-star population \citep{callingham21,yiu24,launhardt22,pritchard21,pritchard24,driessen24}. 
Most radio stars are low-mass M-dwarfs \citep{callingham21,yiu24,pritchard24}, and they are mostly chromospherically active, fast-rotating, young stars, sometimes contained in binary systems such as RS CVn variables. There are several mechanisms responsible for nonthermal radio emission in cool and ultracool (spectral type $\geq$ M7\,V) dwarfs: 
incoherent gyro-synchrotron (or synchrotron) emission probably coming from their magnetosphere, in particular in the form of Jovian-like radiation belts (e.g. \citealt{berger01,berger02,berger06,loinard07,burgasser15,climent23,kao23}); solar-like plasma emission  recently claimed for the first time in another star \citep{callingham25}; and the electron-cyclotron maser (ECM, \citealt{melrose82,treumann06}) seen in magnetic stars, brown dwarfs, and planets (e.g., \citealt{zarka98,route12,route13,williams15,route16,lynch16,kao16,williams17,pineda17,kao18,vedantham20,leto21,rose23,zhang23,bloot24}). Gyro-synchrotron is characterized by a weak-to-moderate circular polarization and is usually more persistent than the bursty, coherent emission of the other two mechanisms. In all cases, the radio emission is related to the presence of magnetic fields and their activity.

The low-mass system 2MASS~J05082729--2101444 (hereafter 2M0508--21) is an intriguing radio source whose radio emission was first characterized in detail by \cite{kaur24}. The system has also been classified as a complex periodic variable (CPVs, also called scallop-shell stars, \citealt{bouma24}), a recently discovered sample of young (10--100 Myr), low-mass stars that exhibit periodic, complex optical light curves with sharp dips ($\sim 5-10\%$ decrease in flux) and additional features \citep[][]{stauffer17,bouma24}. The light-curve shapes slowly change over timescales of hundreds to thousands of cycles, but the periodicity is maintained over years at least. \citet[][]{bouma24} summarized the known sample of 50 confirmed CPVs, plus 13 candidates, many discovered by inspecting the two-minute cadence of Transiting Exoplanet Survey Satellite (TESS) data. Their photometric periodicities are typically a fraction of a day and are thought to track the rapid stellar rotation. CPV optical dips can be explained neither by classical stellar spots---which cannot reproduce the observed sharp features---nor by grazing planetary transits, due to the mid-term variability in depth, shape, and sometimes the phase shift of the dips. Instead, the most convincing scenarios involve the presence of corotating circumstellar material \citep{stauffer17,gunther22}; that is, either gas from the star trapped in huge prominences  \citep[e.g.][]{waugh22,sanderson23} or opaque dust-like material \citep{stauffer17,bouma24}, and it is possibly embedded in a plasma torus \citep{bouma25} or an outgassing rocky planet 
(\citealt[][]{vanlieshout18} and references therein).
CPVs are very rare ($\sim$1\% of the youngest $\sim$1\% M-dwarfs, \citealt{rebull18,gunther22}) and their occurrence  decreases rapidly with age \citep{rebull22}.

Of the 50 high-quality CPVs in the sample of \cite{bouma24}, half are flagged as possible unresolved binaries using indirect indicators, such as the {\em Gaia} astrometric error and the presence of multiple photometric periods. 
If all these suspected binaries are real, the binary rate among the CPVs would be placed at about 50\%. This is noteworthy because, considering that CPVs are predominantly M-dwarfs belonging to young associations and moving groups, we might expect their binary rate to be approximately 30--40\%, the binary fraction of M-type dwarfs measured in nearby young moving groups \citep[][]{shan17}. Therefore, CPVs appear to exhibit an enhanced binary rate compared to the background conatal population (a similar observation was noted by \citealt[][]{stauffer17} for 23 CPVs in Upper Sco). This suggests a possible physical connection between binarity and the CPV phenomenon, either indirectly or directly. A plausible indirect link could be that binarity tends to reduce disk lifetimes, thereby predisposing stars to more rapid rotation during the pre-main-sequence phase \citep[e.g.,][]{kraus12,rebull18}, which is observed to be a characteristic of CPVs \citep[e.g.,][]{stauffer17,stauffer21,bouma24}. Binaries might also play an active role in the mechanism responsible for CPVs, by the stellar components providing each other extra plasma and an additional energy budget in the stellar magnetospheres, or by perturbing the dynamics of the gaseous or solid material in the system.
It has also been proposed that CPV light curves may be explained by close, tidally synchronized interacting binaries, a hypothesis that has not been tested. Despite the obvious utility of characterizing CPVs' binary systems for the purpose of investigating a possible direct relationship between stellar binarity and CPVs, to the best of our knowledge this has yet to be done.

2M0508--21 has been recognized as a close binary star system, whose components were resolved in an archival image from Keck near-infrared camera adaptive optics (NIRC-AO), with a projected separation of about 50 mas (B\'ejar et al. in prep., hereafter Paper II). Given the small angular separation, the system remained unresolved in other existing near-infrared (NIR), optical, and radio observations. One challenge in studying radio emission from close binary systems is that interferometers such as Karl Jansky's Very Large Array (VLA), the upgraded Giant Metrewave Radio Telescope (uGMRT), Low Frequency Array (LOFAR), and the Australian Square Kilometer Array Pathfinder (ASKAP) cannot resolve binary systems at low frequencies ($\lesssim 10$ GHz), if they have subarcsec separation. Therefore, very long baseline interferometry observations become the only option to determine if the radio emission is associated with one or both components.
The precise characterization of the stellar parameters in young binary systems is particularly interesting, because evolutionary models predicting the stellar parameters are poorly tested at these ages, when these stars are still in the pre-main sequence. 
Therefore, it is fundamental to precisely determine the masses of these systems.

2M0508--21 is a photometrically typical CPV. It is one of the very few known to be radio-loud \citep[another being DG~CVn,][]{kaur25}. It is arguably the most studied both in the optical and NIR
(Paper\,II), and in the radio \citep{kaur24,kaur26}, showing complex short- and long-term variability, including flares. We based this study on our three new Very Long Baseline Array (VLBA) observations and one archival Keck Near-Infrared Camera Adaptive Optics (NIRC-AO) observation. We combined the advantages of high-precision radio astrometry of both components with a large set of multi-epoch RV observations obtained with the Calar Alto High-Resolution search for M-dwarfs with Exoearths with NIR and optical \'Echelle Spectrographs \citep[CARMENES,][]{quirrenbach14}. 

The layout of this paper is as follows.  In Section~\ref{sec:source} we discuss the  properties of the binary system. In Section \ref{sec:obs} we present VLBA observations and data analysis. The details of the Keck-NIRC, the CARMENES RV, and photometric data are presented in Paper II. Section \ref{sec:fit} illustrates the fitting procedure. We discuss the results in Section \ref{sec:results}, which are then put in the context of earlier and ongoing radio and optical observations in Section \ref{sec:previous}. We summarize our main findings in Section \ref{sec:conclusions}.

\begin{table*}
\caption{Properties of the VLBA detections of the two components of 2M0508--21AB.}
\scriptsize
\label{table:vlba_data} 
\centering
\begin{tabular}{lccccccc}
\hline
\hline  
\noalign{\smallskip} 
Date &  $\alpha$ (J2000) & $\sigma_{\alpha}$ & $\delta$ (J2000) & $\sigma_{\delta}$ & Flux peak & Flux density &  Flux rms  \\
 & (hh:mm:ss.s) & (s) & ($^{\circ}$:$'$:$''$) & ($''$) & ($\mu$Jy~beam$^{-1}$) & ($\mu$Jy) & ($\mu$Jy~beam$^{-1}$)  \\
\noalign{\smallskip} 
\hline
\noalign{\smallskip} 
  \multicolumn{8}{c}{2M0508--21A}\\ 
\hline
\noalign{\smallskip} 
14 March 2025 & 05:08:27.3568446 & 0.0000015 & $-$21:01:44.693659 & 0.000062 & 510$\pm$16 & 521$\pm$30 & 18 \\
16 April 2025  & 05:08:27.3575719 & 0.0000024 & $-$21:01:44.687780 & 0.000098 & 344$\pm$21 & 341$\pm$36 & 24 \\
07 May 2025  & 05:08:27.3581118 & 0.0000018 & $-$21:01:44.688931 & 0.000065 & 610$\pm$22 & 671$\pm$40 & 26 \\
\noalign{\smallskip} 
\hline
\noalign{\smallskip} 
  \multicolumn{8}{c}{2M0508--21B} \\
\hline
\noalign{\smallskip} 
14 March 2025  & 05:08:27.35518179 & 0.0000020 & $-$21:01:44.693057 & 0.000081 & 398$\pm$17 & 406$\pm$30 & 18 \\
16 April 2025  & 05:08:27.35555258 & 0.0000027 & $-$21:01:44.682834 & 0.000110 & 320$\pm$21 & 338$\pm$38 & 24  \\
07 May 2025  & 05:08:27.35590299 & 0.0000022 & $-$21:01:44.681514 & 0.000071 & 517$\pm$22 & 545$\pm$39 & 26 \\
\noalign{\smallskip} 
\hline
\end{tabular}
\end{table*}

\begin{figure*}[ht]
\centering
\includegraphics[width=16cm]
{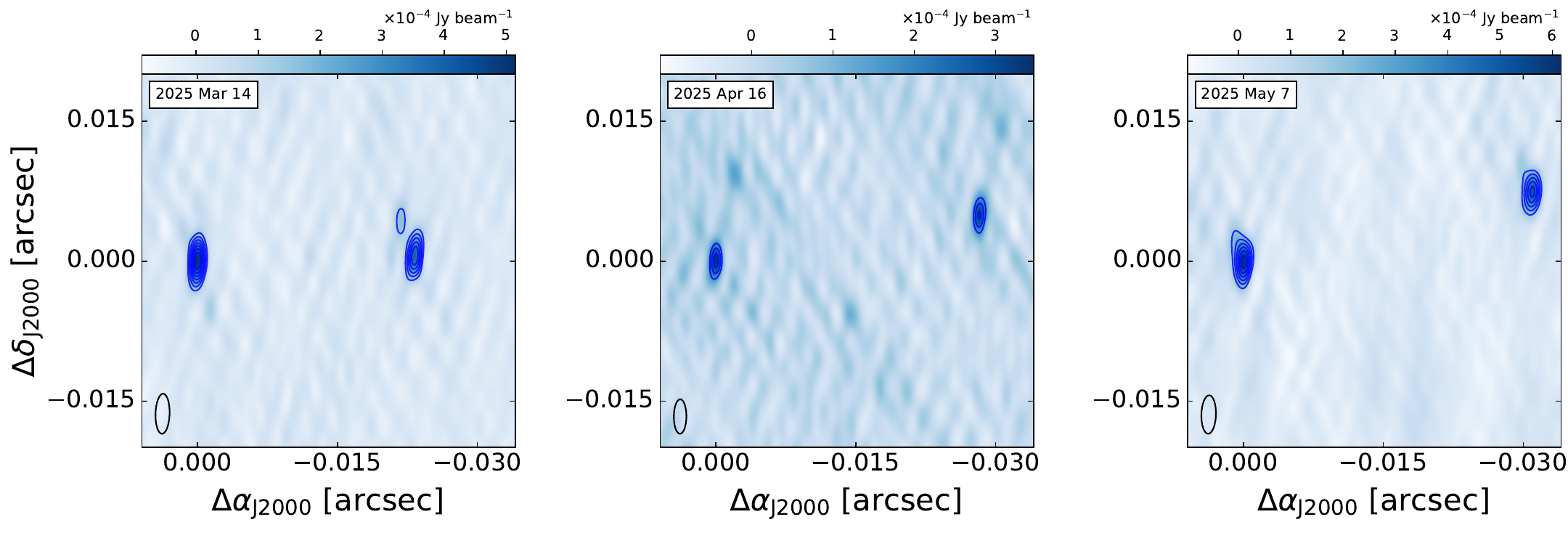}
\includegraphics[width=16cm]
{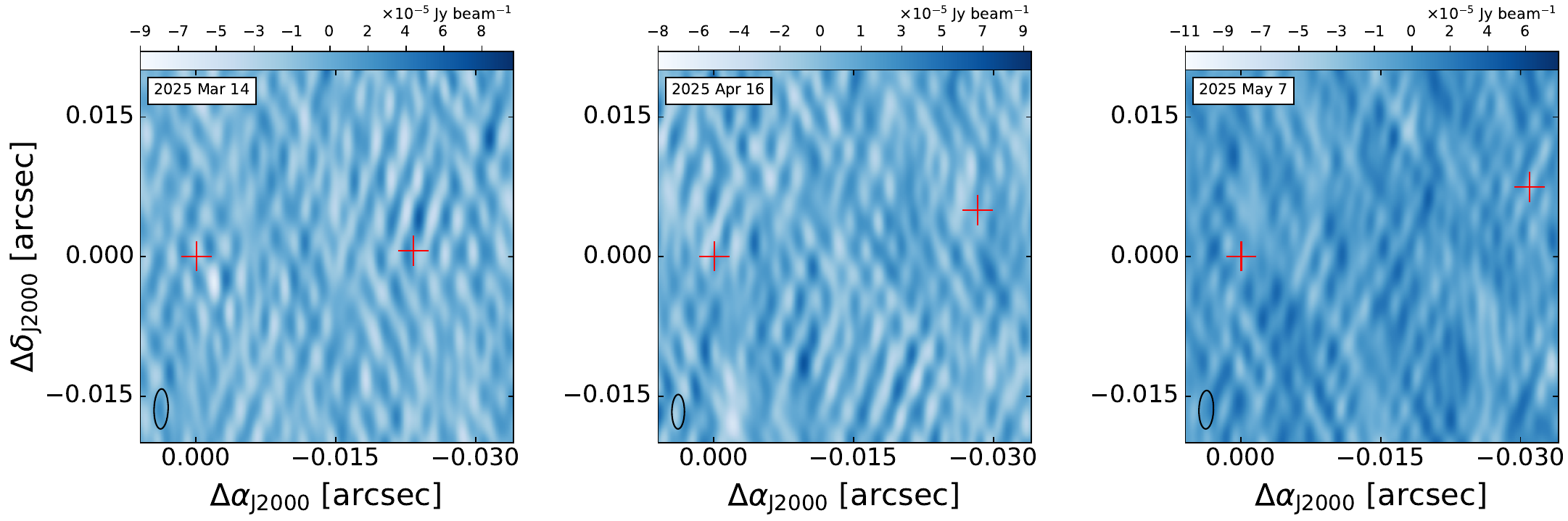}
\caption{Intensity and Stokes V maps of 2M0508--21AB. In each image, 2M0508--21A is on the left and 2M0508--21B on the right. Upper panels: Stokes I maps. In each image, the first contour is at 7$\sigma$ and then in steps of 3$\sigma$, where $\sigma$ is the rms of the images (Table~\ref{table:vlba_data}). Offsets are relative to the position of the 2M0508--21A. Lower panel: Stokes V maps. The rms of the three images are 15, 16, and 15 $\mu$Jy/beam; no emission is observed. The red crosses indicate the position of both stars according to the Stokes I maps. The ellipses in the bottom left corners indicate the sizes of the VLBA-synthesized beams. The date of observation is indicated in the upper left legends.}
\label{fig:vlba_maps}
\end{figure*}

\section{The radio-loud binary 2M0508--21AB}\label{sec:source}

2M0508--21AB, with estimated spectral types of $\sim$M5V \citep{riaz06}, probably belongs to the $\beta$-Pictoris moving group, which would indicate an estimated age of $\sim$10--30\, Myr (Paper II); this is consistent with the observed presence of lithium \citep{miret20}.
A Keck-NIRC AO image obtained in the $H$-band, taken on 8 October, 2012, shows that the system is a visual binary with, at that time, a projected separation of $\sim$50 mas (i.e., 2.5 au, assuming a distance of 48.3 pc, \citealt{gaia21}). The $H$-band image is compatible with two low-mass stars of similar brightnesses, although the relatively large RVs observed in this system indicate that the individual masses are different and that the brightness of the two stars may differ at other wavelengths (Paper II).

The system shows standard CPV optical light curves within the known sample, with TESS-photometry-derived periods of 6.73 and 7.29 h; they were interpreted to track the rotational period of the stars \citep{bouma24}. Photometric observations show persistent periodic dips, including variability in the light-curve morphology and dip depth, which suggests variable density or size (or both) of corotating opaque material around one or both stars. CARMENES RV observations are consistent with a single-line binary system (Paper II). In this compact binary system, the lines are broadened much more than the RV difference, so they are blended in the spectra. 

In recent years, radio-interferometric observations (dedicated or via surveys) with the VLA, uGMRT, and ASKAP have persistently detected the 2M0508--21AB system from 0.5 to 8\,GHz. The flux density is persistently seen to be $\sim$1--4\,mJy, and the radio emission seems to contain a quiescent, mildly variable, and weakly polarized component, ascribable to gyro-synchrotron or synchrotron and possibly rotation-modulated Stokes V burst emission at sub-gigahertz frequencies
\citep{kaur24}. 

The richness of the radio and optical phenomenology could be related to several factors: the binarity, the intrinsic variability of each star, or the presence of opaque material around one or both components. Therefore, a fundamental step toward a full characterization of the system consists of pinpointing the spatial origin of the radio emission and constraining the orbital parameters. This motivated the VLBA observations of this source, which are presented in the following section. 

\section{Observations and data reduction}\label{sec:obs}

Observations of the 2M0508--21AB binary system were obtained with the VLBA in three epochs (BC312A, BC312B, and BC312C), between March 2025 and May 2025 (Table~\ref{table:vlba_data}). The three epochs were observed at 4.85 GHz (Band-C) with a total bandwidth of 512 MHz per polarization and a 4 Gbps recording rate. Observing sessions consisted of switching scans between the target and the phase reference calibrator, J0508$-$2020, which spent approximately 1 minute on the calibrator and 2.5 minutes on the target. The total integration time was 2.5 hours for each epoch. The ICRF position of the phase reference calibrator assumed during the correlation was R.A. = 05:08:47.924512 and  Dec. = $-$20:20:06.44833. 
The fringe finder calibrators J0237$+$2848 and J0457$-$2324 were occasionally observed during each session. The secondary calibrator J0513$-$2159 was observed every 30 minutes and was used to improve astrometric accuracy. Additional 30 minute geodetic-like blocks were observed at the beginning and end of the observing runs.

We reduced the data with the Astronomical Imaging Processing System ({\tt AIPS}; Greisen 2003) following standard procedures for phase-referencing observations \citep[][]{torres07,ortizleon17}  as described in \citet[][]{curiel20,curiel22}. First, corrections for the ionospheric dispersive delays were applied. Then, we corrected for post-correlation updates of the Earth orientation parameters. Corrections were also applied for the digital sampling effects of the correlator. The instrumental single-band delays caused by the VLBA electronics, as well as the bandpass shape corrections, were determined from a single scan on the fringe finder calibrator and then applied to the data. Amplitude calibration was performed using the gain curves and system temperature tables to derive the system equivalent flux density of each antenna. We then applied corrections to the phases for the antenna parallactic angle effects. Multiband delay solutions were obtained from geodetic-like blocks, which were then applied to the data to correct for tropospheric and clock errors. The final step consisted of removing global frequency- and time-dependent residual phase errors obtained by fringe fitting the phase-calibrator data, assuming a point-source model. In order to take into account the non-point-like structure of the calibrator, this final step was repeated using a self-calibrated image of the calibrator as a source model. Finally, calibration tables were applied to the data, and target images were produced using the {\tt CLEAN} algorithm  \citep{clark80}. We used a pixel size of 50 $\mu$as and a pure natural weighting.

The images of 2M0508--21AB are presented in Fig.~\ref{fig:vlba_maps}. The synthesized beam in these images is, on average, 4.0$\times$1.5 mas, with PA = $-$1.27 deg.
Notably, in all observations we clearly detected both components of the binary system, with a projected separation increasing from 23 to 30 mas, in the 54 day time span of the observations, and a counterclockwise motion (Fig.~\ref{fig:vlba_maps}). Both components have a similar and mildly variable radio flux density of between $\sim$0.34 and 0.67 mJy (Table~\ref{table:vlba_data}).
Since both stars are very similar in terms of near-infrared and radio brightness, we cannot identify which component is the most massive. Hereafter, we define 2M0508--21A as the primary star and 2M0508--21B as the secondary star, located in the left and right of Fig.~\ref{fig:vlba_maps}, respectively.

The positions of the two components were obtained with Gaussian fits to the brightness distribution (Table~\ref{table:vlba_data}). 
The two radio sources appear unresolved in the three observed epochs. The only possible exception is that 2M0508--21A appears to be marginally resolved in the third epoch with a deconvolved size of 1.24$\times$0.52 mas and PA = 3.57 degrees.
In Table~\ref{table:vlba_data}, we provide the fitted positions together with their associated uncertainties, as well as the peak and integrated flux densities obtained from these fits. 

We also obtained 119 RV measurements using the two-arm CARMENES spectrograph in the visible (VIS) and NIR channels 
\citep{quirrenbach16, quirrenbach18}. 
The final dataset consists of 60 RV measurements with CARMENES VIS and 59 with CARMENES NIR. The spectra were reduced using the {\tt CARACAL} pipeline \citep{caballero16,trifonov18}, and the RVs were obtained using the {\tt serval}\footnote{\url{https://github.com/mzechmeister/serval}} pipeline \citep{zechmeister18}. The average signal-to-noise (S/N) 
per pixel of the spectra ranges is from 20 to 50 at 745\,nm and from 20 to 80 at 1221\,nm, and the median error in RV is 70\,m/s in the VIS channel and 220\,m/s in the NIR channel. Further information on these data are provided in Paper\,II.

The three radio observations were obtained over a time span of approximately two months. Therefore, these astrometric observations alone do not provide sufficient information about the proper motion and parallax of the system, nor about the orbital motion and masses of its two components. However, by combining these astrometric data, the Keck-NIRC data, and the CARMENES optical/infrared RV data, we were able to fit the orbital motion of the binary system and obtain its total dynamical mass.

\begin{figure}[!t]
\centering
\includegraphics[width=0.47\textwidth] {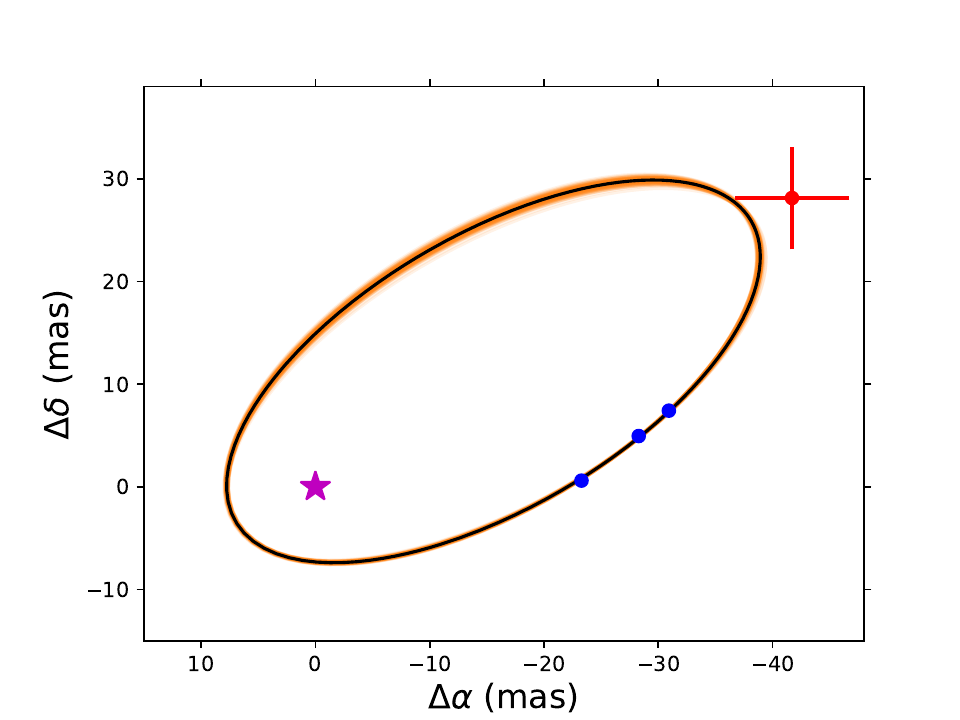}
\includegraphics[width=0.47\textwidth]{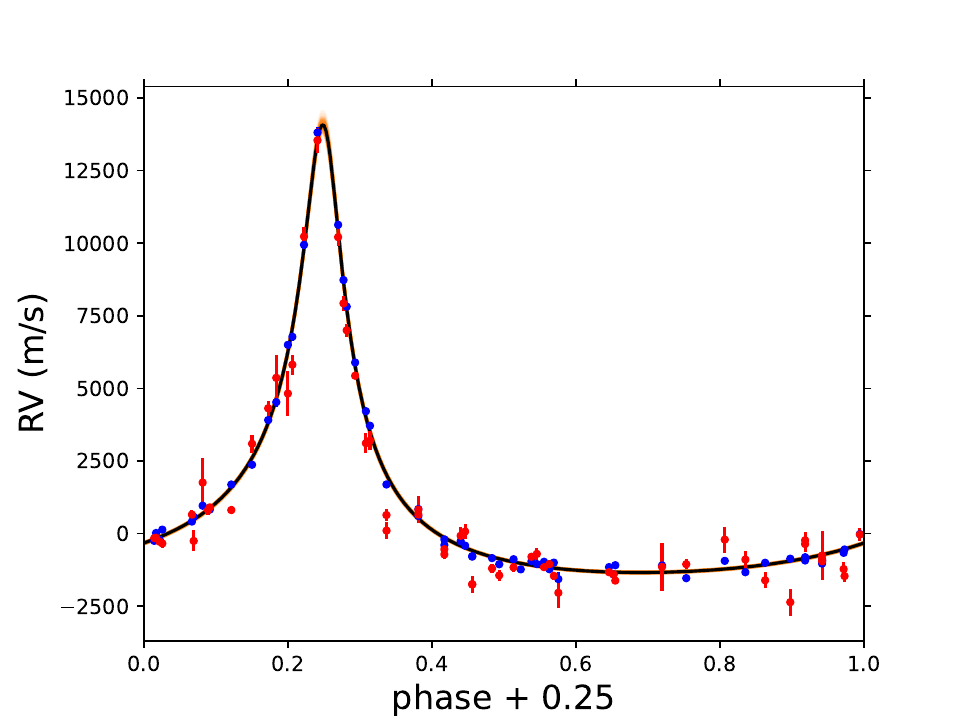}
\caption{Combined fit solution. The top panel shows the fitted solution for the relative astrometry of 2M0508--21B around 2M0508--21A (marked with a magenta star). VLBA epochs are shown in blue, and the Keck NIRC epoch is shown with a red cross. The error bars of the VLBA data are smaller than the size of the blue symbol. The binary system rotates counterclockwise. The bottom panel shows the fit solution for the single-line RV data. The optical RV data are shown in blue, and the NIR RV data are shown in red. The solid black line shows the best-fit solution for both relative astrometry and RV. A total of 100 random solutions obtained with the {\tt MCMC} code are shown in orange.}
\label{fig:combined_fit}
\end{figure}

\begin{table}
\caption{Combined astrometric fit.}
\label{table:fitted_parameters} 
\centering
\begin{tabular}{lr}
\hline\hline  
\noalign{\smallskip} 
Parameter & Fitted parameter \\
\hline
\noalign{\smallskip} 
$P$ (d)             & 801.40$\pm$0.30  \\
$T_{0}$ (JD)          & 2460643.79$\pm$0.23  \\
$e$                        & 0.7140$\pm$0.0026  \\
$\omega$ (deg)    & 5.23$\pm$0.21  \\
$\Omega$ (deg)    & 122.24$\pm$0.37   \\
$a$ (mas)              & 26.964$\pm$0.087  \\
$i$ (deg)                & 46.75$\pm$0.69   \\
$\gamma$  (km s$^{-1}$)  & 0.8893$\pm$0.0093   \\
$K$ (km s$^{-1}$)\tablefootmark{a} & 7.712$\pm$0.074   \\
$\chi^{2}$, $\chi^{2}_{\rm red}$\tablefootmark{b} & 852.30, 7.22 \\
\noalign{\smallskip} 
\hline
\noalign{\smallskip} 
  \multicolumn{2}{c}{Other parameters} \\
\noalign{\smallskip} 
\hline
\noalign{\smallskip} 
$\pi$ (mas) & 20.703 (fixed) \\
$M_{\rm AB}$ (M$_{\odot}$) & 0.4591$\pm$0.0073  \\
$a$ (au)       & 1.303$\pm$0.012 \\
\hline
\end{tabular}
\tablefoot{
\tablefoottext{a}{$K$ is the brightness-weighted difference between the $K$ values of the individual components.}
\tablefoottext{b}{$\chi^{2}$ and $\chi^{2}_{\rm red}$  of the astrometric fit. The residuals of the RV dominate the residuals of the fit.}
}
\end{table}

\section{Analysis}\label{sec:fit}

In order to obtain the orbital motion of the binary system, we followed the fitting procedure presented by \cite{curiel24}. We used a Markov chain Monte Carlo ({\tt MCMC}) code that fits relative astrometry and RV simultaneously. Here, we used the open-source package {\tt lmfit} \citep[][]{newville20}, which uses a nonlinear least-squares minimization algorithm to find the best fit of the observed data.
This Python package includes Levenberg–Marquardt minimization and {\tt emcee} \citep[][]{foremanmackey13}.
The code we used in this work includes the possibility of adding RV data in the astrometric fitting. 
This combined fit removes the ambiguity in the position angle of the ascending node ($\Omega$ and $\Omega$ $+$ 180 deg).

When fitting the combined astrometric and RV data, we weighted the data by the positional errors of both relative coordinates ($\Delta\alpha$ and $\Delta\delta$) and RV errors (Paper\,II). We used 250 walkers and ran the {\tt MCMC} for 10\,000 steps with a 700 step burn-in, at which point the chain length is over 50 times the integrated autocorrelation time. We minimized the function $\chi^{2}$ to obtain a maximum-likelihood estimate of the model parameters that are being fitted  \citep[][]{curiel24}.

The measurements consist of a total of three VLBA epochs of absolute radio astrometry spanning 54 days, 119 RV measurements of CARMENES single-line RV data spanning 8.1 years, and one epoch of relative astrometry extracted from the archive image obtained with the Keck telescope (Paper\,II for more details). Since we did not have enough absolute astrometric observations to fit the proper motion and parallax of the system, we used the precise relative astrometric measurements of the system and the RV data to obtain a combined fit of the relative astrometry with the RV data. We used a parallax value of 20.703$\pm$0.059 mas, as reported by {\em Gaia} DR3 for this system \citep[][]{gaia23,babusiaux23}, to obtain the total mass of the binary system.

The combined astrometric model  includes the seven parameters needed to fit the orbital motion of the binary system and two additional parameters, one for the relative velocity amplitude between the primary and the secondary ($K$), and the other for the barycentric RV of the binary system ($\gamma$).
However, since RVs are single lined (Paper II),
$K$ is the brightness-weighted difference between the K values of the individual components and $\gamma$ is also weighted by the brightness of the two components.
The seven orbital parameters are the orbital period ($P$), time of the periastron passage ($T_{0}$), eccentricity ($e$), longitude of the periastron ($\omega$), position angle of the ascending node ($\Omega$), semimajor axis of the secondary's orbit around the primary ($a$), and inclination angle of the orbital plane ($i$). 
The four common parameters between astrometry and RV are $P$, $T_{0}$, $e$, and $\omega$.

\begin{figure*}[ht]
\centering
\includegraphics[width=15cm]
{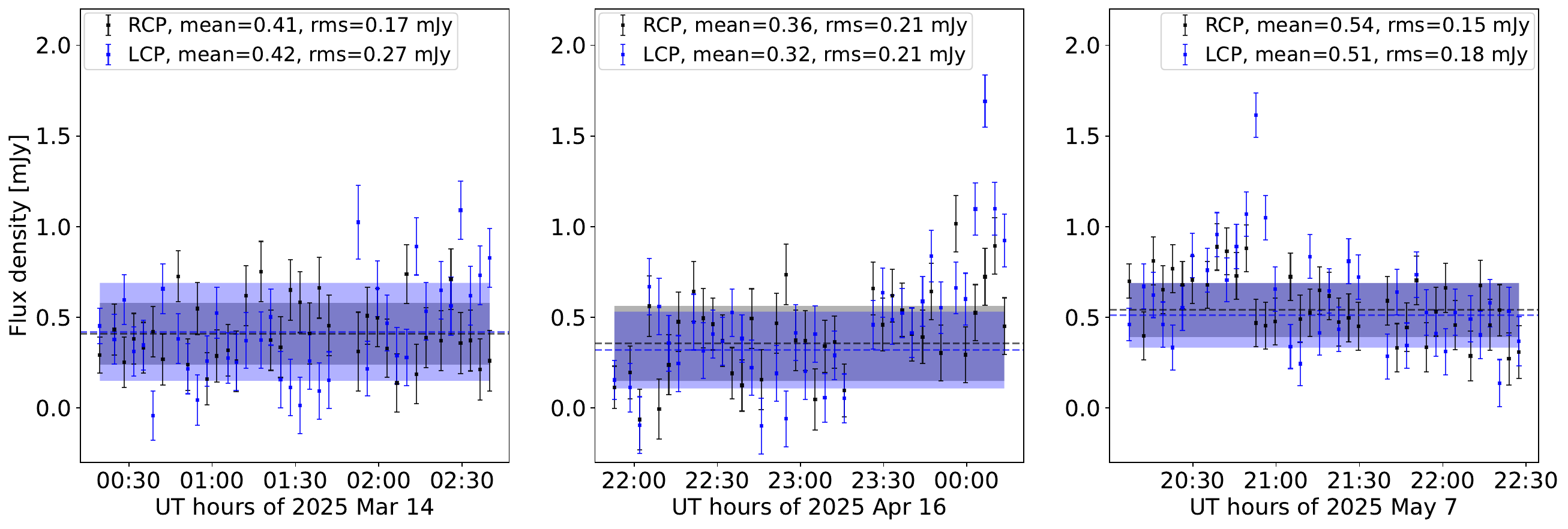}
\includegraphics[width=15cm]
{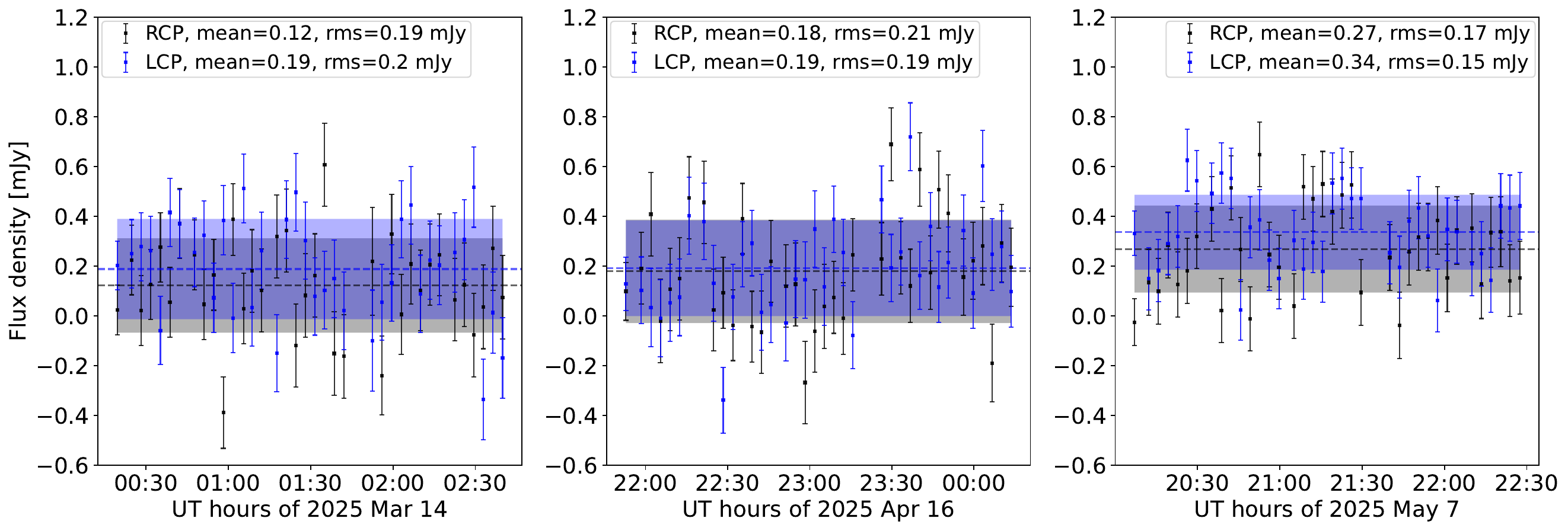}
\caption{Upper panels: Radio light curves of left circularly polarized (LCP) and right circularly polarized (RCP) emission; these were computed as the real part from the visibility plane with the phase center at the position of 2M0508--21A. Bottom panels: Same as the upper panels but with the phase center at the position of 2M0508--21B. The three VLBA epochs are shown from left to right. For each epoch and polarization, the horizontal lines indicate the mean flux densities, and the shadow color bands show the $\pm$1$\sigma$ noise level for both polarizations.}
\label{fig:vlba_light_curve}
\end{figure*}

\section{Results}\label{sec:results}

The results of this combined fit are shown in Fig. \ref{fig:combined_fit} and summarized in Table~\ref{table:fitted_parameters}.
We find that the orbit of the binary system is eccentric ($e = 0.7140\pm0.0026$), with a semimajor axis of 26.964$\pm$0.087 mas ($a$ = 1.30 au) and an orbital period of 2.19 years. The inclination angle of the binary system is 46.75$\pm$0.69 degrees, which indicates that the orbit is prograde. 
From the combined fit, we find that the total mass of the binary system is 0.4591$\pm$0.0073\,M$_{\odot}$. Since we used a fixed parallax, the error of the total mass is overestimated. 
In addition, the fit results show that the velocity amplitude is 7.7 km~s$^{-1}$.

Figure~\ref{fig:combined_fit} shows that the high-precision VLBA astrometric data constrain the combined fit more than the less precise Keck relative astrometric data.
Fig.~\ref{fig:combined_fit} includes (in color) 100 random solutions obtained from the combined fit. It shows that the dispersion of the orbital+RV solutions is very small, which is consistent with the small errors of the fitted parameters in Table \ref{table:fitted_parameters}. Fig.~\ref{fig:corner} shows the correlation between the fitted parameters. This figure also shows a small dispersion in the fitted
parameters. The residuals obtained with the {\tt MCMC} code are small and well behaved. The reduced $\chi^{2}$ value of 7.22 reflects the relatively small residuals of the fit of the RV data.

We only used four relative astrometric epochs, which cover a small section ($\sim$20\%) of the binary system's orbit (Fig.~\ref{fig:combined_fit}). 
Obtaining a good fit of the combined data (Table \ref{table:fitted_parameters})
is mainly due to the time coverage of the RV data (8.1 years; i.e., more than three orbital periods) and the high precision of the absolute astrometry obtained with the VLBA.
In addition, the RV data restrict several of the fitted parameters ($P$, $e$, $\omega$, $T_{0}$, and $K$), while the relative astrometric data restrict the other fitted parameters, giving a good orbit fit.
Additional VLBA observations covering most of the binary system's orbit will help to improve the combined astrometry and RV fit and thus to improve the values of the fitted parameters. Furthermore, such absolute astrometric data would make it possible to simultaneously fit the proper motions and parallax of the binary system's barycenter and the orbital motion of both stars around their center of mass, providing the masses of the binary system and the individual stars. 

Figure~\ref{fig:vlba_maps} shows that the data obtained with the VLBA do not show Stokes V flux emission above the root mean square (rms) of the maps (3$\sigma$ = 45, 48, and 45 $\mu$Jy\,beam$^{-1}$) at the position of either stellar component, in any of the observed epochs, which implies an upper limit on the circular polarization fraction V/I $\lesssim$ 10\%. 
Both stars exhibit a flux-density variability of approximately 0.2--0.3\,mJy between consecutive observations (Table \ref{table:vlba_data}). The combined flux density of the two stars (total flux = 0.93, 0.68, and 1.22 mJy) indicates a variation of approximately $-37\%$ and $+80\%$ between consecutive observations and a mean flux density of 0.94\,mJy, with a standard dispersion of 0.27\,mJy.

Figure~\ref{fig:vlba_light_curve} shows the radio light curves obtained from the visibility plane with the phase center at the position of both the primary star (top panels) and the secondary star (bottom panels). For each epoch and each polarization, we computed the rms of the data. The mean flux density  and  rms values are provided in the legends of Fig.~\ref{fig:vlba_light_curve}. This figure shows no apparent flare events within the $\sim$2.5\,h of each observation. 
However, the second and third epochs show an apparent temporal flux-density variation, mainly from 2M0508--21A. In addition, in the second and third epochs, the left circularly polarized (LCP) emission reaches fluxes of $\sim$3$\sigma$ at about 00:15 UT and 20:50 UT, respectively.  
These tentative polarization events occur over very short periods of time. The radio light curves of the secondary star do not show similar polarization events. These results suggest that short-lived polarized transient emission may be associated with 2M0508--21A.
An inspection of the individual radio light curves of the two components, obtained from the image plane, is shown in App. \ref{app:individual_lc}. Figs.\,\ref{fig:flux_10min} and \ref{fig:flux_2min} show Stokes~I and Stokes~V as functions of time, respectively.
They do not show relevant bursts or clear trends in time within each observation, although there are hints of short-lived variability, which is difficult to assess because of the low S/N with short-term integrations of 10 and 2 min. 
Stokes~V also shows no apparent polarization in the radio emission, which is probably also due to the low S/N of the short integration time intervals used to search for polarization.
Additional VLBA observations are needed to confirm the apparent short-lived polarized transient emission seen in the radio light curves presented in Fig.~\ref{fig:vlba_light_curve}.
In summary, the radio emission generally appears to be quiescent in the three VLBA 4.85 GHz observed epochs, with no flares or detectable circular polarization, except for marginal evidence of transient polarization from 2M0508--21A. 

\begin{figure*}[ht]
\centering
\includegraphics[width=.45\textwidth]{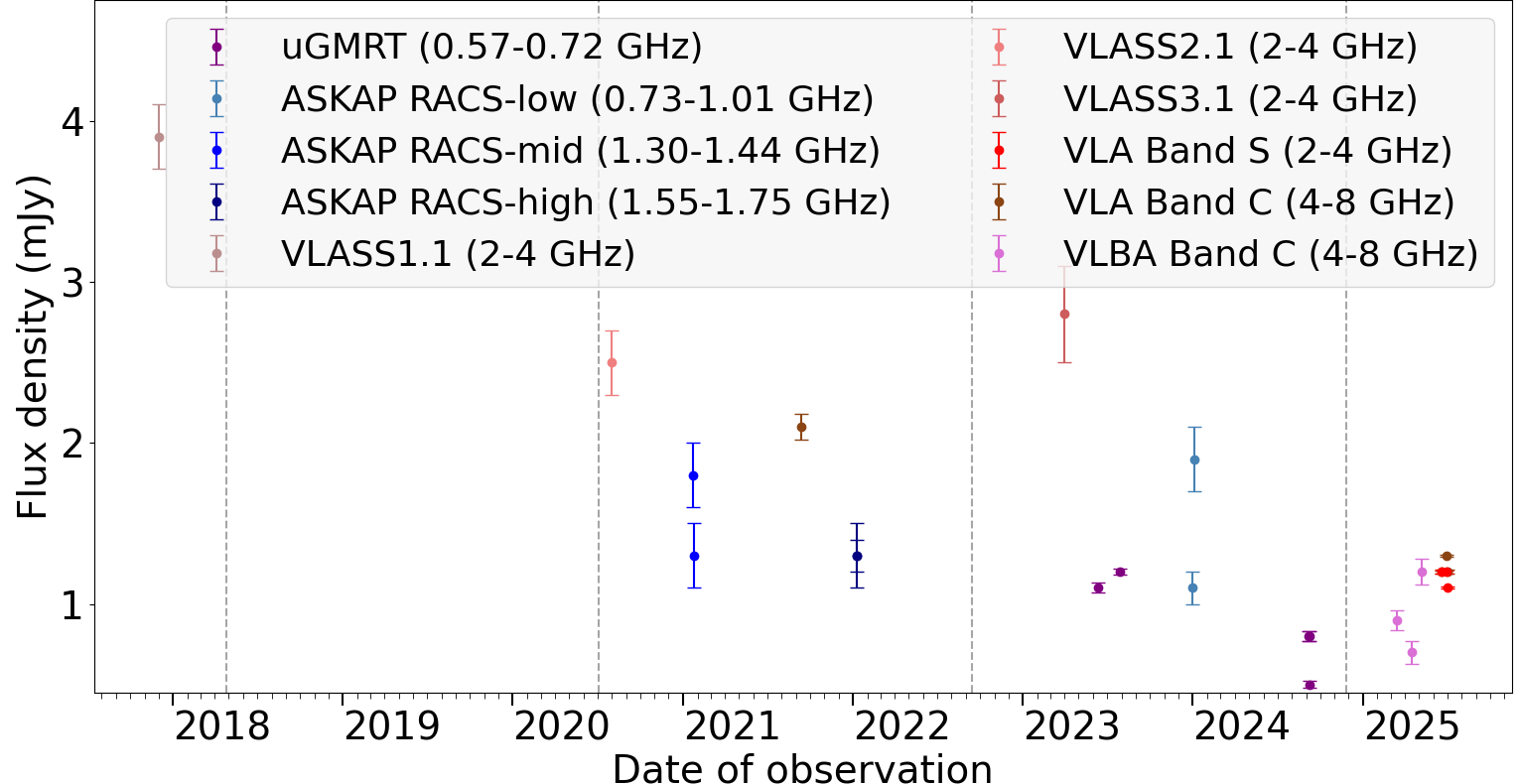}
\includegraphics[width=.45\textwidth]{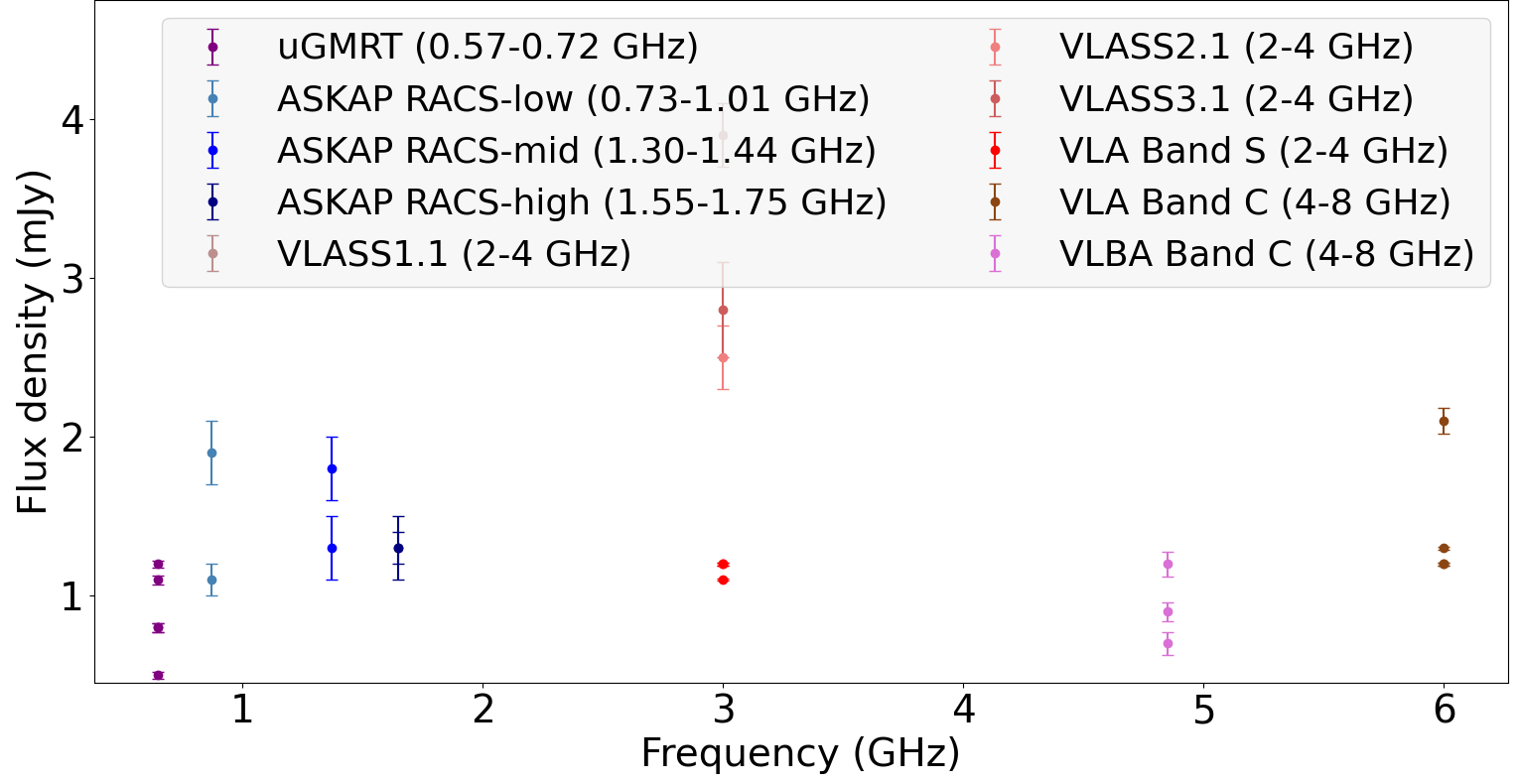}
\caption{Measured flux density of all known radio observations of J0508$-$21AB from surveys and dedicated observations \citep{kaur24,kaur26}, including those presented here, as a function of observing date (left panel) and frequency band (right panel). The vertical dashed lines on the left panel show the periastron passages according to the 801 day solution. For the VLBA data, we combined the flux-density values of both stars shown in Table \ref{table:vlba_data}, while for the other radio telescopes the source is not resolved.}
\label{fig:flux_radio}
\end{figure*}

\section{Discussion}\label{sec:previous}

\subsection{Long-term flux-density variability}

Figure~\ref{fig:flux_radio} shows the Stokes I flux for all the available observations to date, including surveys carried out with VLASS and ASKAP and those obtained with VLA, uGMRT, and VLBA. 
This figure shows a flux-density variability at all observed frequencies.
This figure also shows a nearly flat spectrum from 0.75 to 6 GHz, with a significant dispersion. However, three observations show a large excess in the flux density at 3 GHz, which could be related to strong outbursts in the source occurring at those epochs. A detailed analysis of the flux-density time variability of the source is beyond the scope of this paper, but it can be found in \cite{kaur24,kaur26}.

\subsection{Origin of the radio emission}

Quiescent radio emission is usually interpreted as gyro-synchrotron or synchrotron radiation produced by mild or ultra-relativistic electrons spiraling in magnetic fields, originating in the corona or radiation belts (or both).
The lack of circular polarization in the VLBA observations and the persistence of quiescent emission without flare events suggest a gyro-synchrotron origin for radio emission \citep[e.g.][]{callingham24}. This nonthermal radio emission is produced by mildly relativistic electrons in the magnetosphere of the star.
The associated brightness temperature is \citep[][]{williams13}

\begin{equation}
\begin{split}
T_{b} &= \frac{c^{2}}{2 k_{B} \nu^{2}} \frac{S_\nu}{\Omega}  \sim \left( 10^{6.5} {\rm K} \right) \left(\frac{\nu}{{\rm GHz}} \right)^{-2} \left( \frac{S_\nu}{{\rm \mu Jy}} \right)  \left( \frac{d}{{\rm pc}} \right)^{2} \left( \frac{r}{R_{\rm Jup}} \right)^{-2} \\  
&\sim  \left( 3.14 \times 10^{8} K \right) \left( \frac{S_\nu}{{\rm \mu Jy}} \right)  \left( \frac{r}{R_{\rm Jup}} \right)^{-2}
\end{split}
,\end{equation}

\noindent
where S$_{\nu}$ is the flux density of the star (between 340 and 670\,$\mu$Jy, Table\,\ref{table:vlba_data}), $\nu$ is the frequency, $\Omega$ is the solid angle of the radio-emitting region, $d$ is the distance to the source, and $r$ is the effective radius of the emitting region in Jupiter radii (R$_{\rm Jup}$). 
Sources with $T_{b} \gtrsim 10^{12}$ K are considered coherent, whereas sources with $T_{b} \lesssim 10^{12}$ K are considered incoherent \citep[][]{dulk85,seaquist93}.
Since the emission of the source is not spatially resolved, and we do not see any clear short time variability that could place an upper limit on the emitting size, we have no constraints on the emitting radius. Therefore, we can only make educated guesses.
Using the radius of the stars ($r$ = 4.3\,R$_{\rm Jup}$) obtained in Paper~II, the brightness temperature of both stars is $T_{b} \sim$ 0.6--1 $\times 10^{10}$\,K, which is consistent with gyro-synchrotron and synchrotron emission.
On the other hand, length scales of $r \lesssim$ 0.33\,R$_{\rm Jup}$ would give brightness temperatures of $T_{b} \gtrsim$ 1--2 $\times 10^{12}$\,K, which  would be problematic for this interpretation due to the brightness temperature constraints of gyro-synchrotron emission \citep[e.g.,][]{callingham24}, and the radio emission could be associated with electron-cyclotron maser (ECM) emission. However, in this case, the radio emission would be highly polarized ($\gtrsim$ 90\%), which is not observed in the radio emission of these stars. Therefore, if the radio emission from the VLBA data is extended ($r >$ 0.33\ R$_{\rm Jup}$), although unresolved in the present observations, its would be of gyro-synchrotron or synchrotron origin. 
2M0508--21A was marginally resolved in one of the three observed epochs (Section\,\ref{sec:obs}), with a deconvolved size (1.24\,mas versus 0.52\,mas, with an average size of 0.88\,mas) that would correspond to a physical radius of $\sim$88.9\,R$_{\rm Jup}$.
If we use this physical size as the radio-emitting region, we obtain a brightness temperature of $T_{b} \sim$ 1--3 $\times 10^{7}$\,K. In this case, the radio emission could be synchrotron in nature, probably produced in an extended radiation belt around the star; such radiation belt would be similar to, but much more extended than, that found in the ultra-cool dwarf LSRJ1835$+$3259 \citep[][]{kao23,climent23}. 
However, for synchrotron emission by ultra-relativistic electrons, the expected brightness temperature is $T_{b}\, \gtrsim 10^{10}$\,K, which is much higher than the estimated brightness temperature, thus indicating mildly relativistic electrons.
In summary, in the absence of clear constraints on the emitting region, the radio emission at 4.85\,GHz appears to be consistent with gyro-synchrotron or synchrotron radiation. Further observations could shed more light on the mechanism that produces the observed radio emission; the possible mechanisms are discussed in dedicated multiwavelength studies \citep{kaur24,kaur26}.

\subsection{Comparison with other results}

2M0508--21AB is currently one of the best-characterized CPVs in radio among the known samples \citep{bouma24}. The only comparable radio-observed case from the same sample is DG CVn, which is a similar binary in terms of age, separation, photometry, and persistent transient components observed at radio frequencies \citep[][]{kaur25}. In both systems, the variability and features in both radio and optical light curves show a complexity that leaves many open questions. These include whether the long-term variability (weeks, months) of the radio flux densities and optical light-curve morphology are related to the sources being binary systems, whether the low-frequency ECM bursts happen at the same primary and/or secondary
rotational phases, whether the optical dips are related to the binary orbit, and which of the two stars is responsible for the radio emission. The study we present here can answer the latter question for the case of 2M0508--21AB, reinforcing the idea that both stars are similar in terms of rotation periods and radio emission and suggesting that both stars may have a similar magnetic field.

This binary is the first CPV with the direct detection of both stellar components with the VLBA and with a full orbital solution, together with DG CVn (P. Boven, priv. comm.). It is also one of the few M-dwarf binary systems observed with multiple radio-emitting components. 2M0508--21AB is only the second M-dwarf system investigated with VLBI after the much older (950\,Myr) M-dwarf binary system GJ~896AB \citep{parsamyan95}, where both components were found to be radio emitters \citep{curiel22}.   
Two older and lower mass binary systems were also observed with VLBI and were found to be radio emitters. In the case of the M7 LSPM~J1314$+$1320AB  binary system (80.8 $\pm$ 2.5\,Myr), only one component was found to be a radio emitter \citep[][]{dupuy16}, while both components in the M8$+$M9 LP~349$-$25AB binary system (262\,Myr) were found to be radio emitters \citep[][]{curiel24}. 
These results show the potential of long-baseline radio observations to resolve compact binary systems and to directly study the individual stars. This may help to understand, for example, the nature of the dips observed in optical light curves of CPV stars.

The three VLBA observations were obtained with different time intervals between consecutive observations. 
The on-source time of each VLBA observation is about 2.5 h, which covers 37\% of the 6.73 h primary rotational period, or 34\% of the 7.29 h secondary rotational period.
Taking into account the primary period, which aligns very well with the dips, the times of the VLBA observations, and the phase of the dips seen in the closest-in-time optical observations (Paper\,II), we find that only the first and third VLBA epochs cover a fraction of the dips ($\sim$50\%). In addition, we estimate that the combined VLBA observations cover about 78\% of the primary rotation period, and about 60\% of the secondary rotation period. Finally, taking into account the radio light curves (Fig.~\ref{fig:vlba_light_curve}) and the phase and time coverage of the optical dips, we find that the processes responsible for the optical dips do not seem to affect the radio continuum, polarization, and radio flares in our VLBA observations at 4.85\,GHz. However, previous observations have shown a possible optical-radio interconnection at 
sub-gigahertz
frequencies \cite[][]{kaur24}. These sub-gigahertz observations also showed clear circular polarization, with a relatively high circular polarization fraction, and highly circularly polarized bursts. Therefore, there appears to be a general trend toward a lower polarization at higher frequencies. 
Due to the time variability of the radio emission observed at low and high frequencies, simultaneous or contemporary observations in a wide range of frequencies are needed to disentangle the frequency and time variability of the intensity and the polarization and a possible correlation between radio emission and optical dips.
In addition, CARMENES data allow the presence of very close binaries in the system with orbital periods shorter than a few tens of days to be discarded, because in these cases the RV difference of both components would be above 30 km~s$^{-1}$ and the spectral lines would be resolved.

\subsection{Comparison with theoretical evolutionary models}\label{sec:models}

In the companion article, Paper\,II, the luminosity of the individual components was determined under the assumption that the system is a binary of equal brightness. This is supported by the $H$-band image from Keck NIRC-AO, which resolved the system in two components of similar brightness with an accuracy of 10\%,
although the two stars probably have different masses and different brightnesses at other wavelengths (Paper II).
In that article, individual masses of 0.188$^{+0.050}_{- 0.060}\,M_\odot$ were estimated by comparing individual luminosity with theoretical evolutionary models of \cite{baraffe15} for the most likely age of the system (10--30\,Myr), considering its membership in the young  $\beta$-Pictoris moving group \cite[][Paper II]{schneider19}. Therefore, we can compare the total mass of the system of  0.459$\pm$0.007 \,M$_\odot$, determined by the combined astrometric and spectroscopic fit in Section\,\ref{sec:fit}, with the sum of the individual masses predicted by  the theoretical evolutionary models (0.38$^{+0.10}_{-0.12}\,M_\odot$). The last value is slightly lower but is in agreement with the total mass obtained with the combined fit. 

\begin{figure}[t]
\centering
\includegraphics[width=0.47\textwidth]{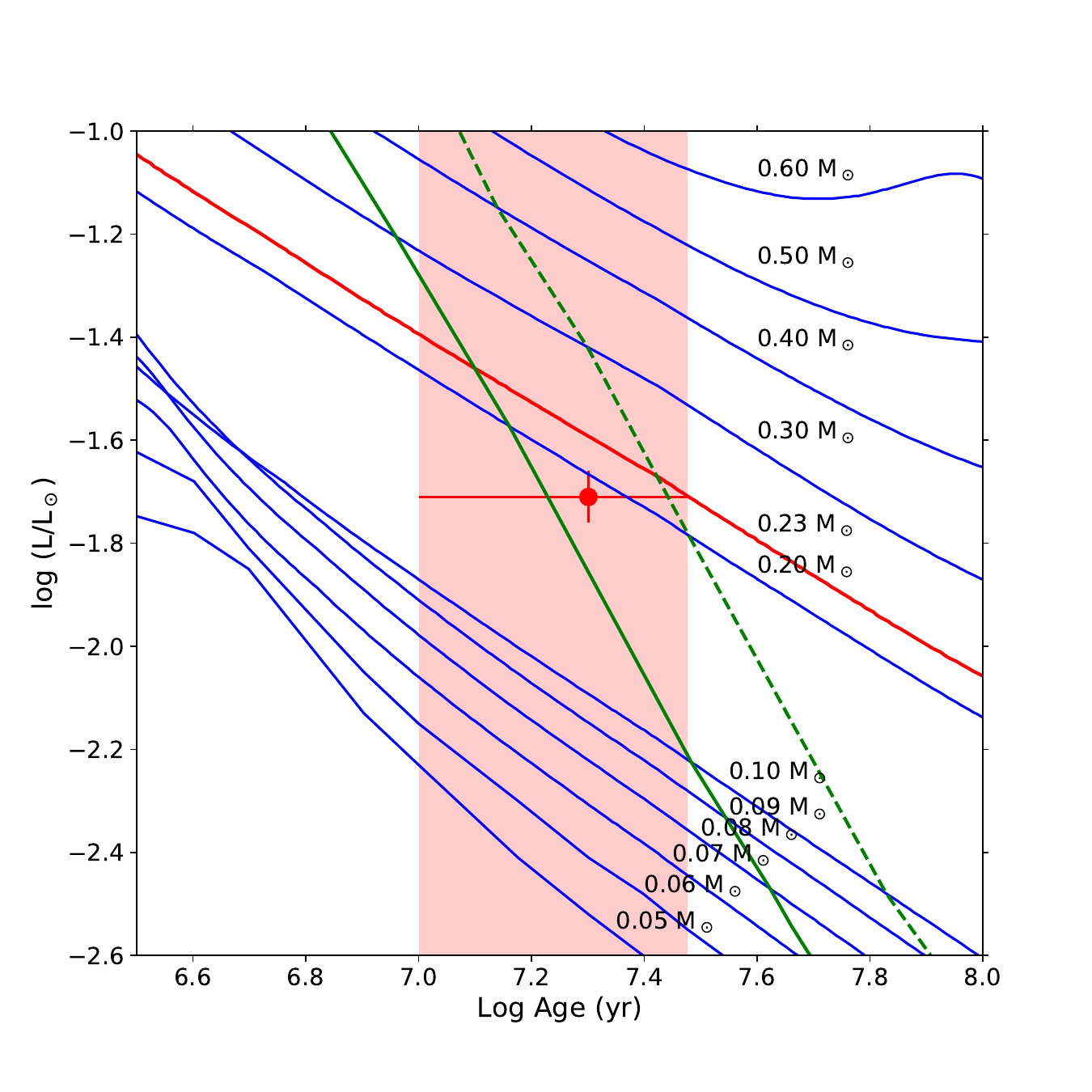}
\caption{Luminosity versus age diagram. The luminosity of individual components (assuming equal brightness) is indicated by a solid red circle, with the vertical error bar indicating its uncertainty. The horizontal error bar and the pink shaded area indicate the estimated age range of the $\beta$-Pictoris moving group. The theoretical evolutionary models of \cite{baraffe15} are represented by solid blue lines, which are labeled with their corresponding masses; the 0.23\,M$_\odot$ case (corresponding to half of the estimated total mass of the system) is shown in red. 
The solid and dashed green lines indicate the 1\% and 99\% boundaries of lithium depletion as a function of mass, respectively.}
\label{fig:E_models}
\end{figure}

In Fig. \ref{fig:E_models}, we plot the evolution of luminosity with age for low-mass stars and brown dwarfs. We include the luminosity of individual components and compare it with the isochrones  for different masses of \cite{baraffe15}. The model for the expected individual mass of an equal-mass binary of 0.23\,M$_\odot$ is indicated in red. We also included the lithium (Li) depletion boundary (1 and 99\% depletion). The Li pseudo-equivalent width (pEW) of this object is measured as pEW(Li) = 0.6$\pm$0.1\,{\AA} from the CARMENES template (Paper II), which is consistent  with complete preservation of Li (or very small depletion of it). From this figure, we can observe that the individual luminosity and mass could only be consistent within 1$\sigma$ with the models in the oldest age range of $\beta$-Pictoris, but in this case a significant Li depletion is expected, so they are not consistent. They could be consistent within 2$\sigma$ with the models of the average age of $\beta$-Pictoris, but we still expect a significant Li depletion; they could be consistent within 3$\sigma$ with Li preservation, but only for the youngest age range of $\beta$-Pictoris.  

Therefore, although both observations and the models could be consistent within $3\sigma$, some tension remains. Current evolutionary models that include magnetic fields, such as SPOTS \citep{somers20}, predict slightly higher luminosities for young M-dwarfs compared to standard nonmagnetic models and therefore cannot explain the differences found.
This tension would disappear if the true total mass were lower ($M_{\rm tot} <$0.40$-$0.30\,M$_\odot$) and, therefore, the individual masses were below $\sim$0.20--0.15\,M$_\odot$. The total mass of the system strongly depends on the distance, and this could be closer if the \textit{Gaia} DR3 parallax is affected by binarity. 
The parallax measured by \textit{Gaia} is of 20.70$\pm$0.06\,mas (distance of 48.3$\pm$2.2\,pc), and a slightly larger value (by 5\,mas) due to orbital motion would imply a distance of 39\,pc and a total mass of 0.24\,M$_\odot$. This would be more consistent with other dynamical mass measurements of $\beta$-Pictoris members of similar spectral types,  such as GJ~3076 (an M5+M6 system with a tentative total mass of 0.10$-$0.26\,M$_\odot$ due to large parallax uncertainties; \citealt{calissendorff22}), or TWA~22AB (a very similar M5+M5.5 system with a total mass of 0.18$\pm$0.02\,M$_\odot$; \citealt{rodet18}). 

On the other hand, as it is pointed out in Paper\,II, to measure such relatively large RV variations from the combined lines of both components, the two components must have different masses, or different masses and brightnesses. 
To resolve these discrepancies, the parallax and mass ratio of the system, and therefore the individual masses of both components, need to be determined.  

\subsection{Periastron time}

Given the large eccentricity of this binary system ($e = 0.71$), the gravitational interaction between the two stars will increase substantially close to the periastron location, possibly producing large perturbations in the corotating material that seems to surround the stars.  
It is not clear whether the closest proximity of the two stars ($\sim$0.37\,au) at the time of periastron would be enough to  significantly increase the activity of both stars. 
Using the combined astrometric solution, we find that the previous two periastron passage times were on 19 September, 2022 and 29 November, 2024, and the next periastron passage will be on 8 February, 2027. The last periastron passage occurred about four months before the first VLBA observation (Fig.~\ref{fig:flux_radio} and Table~\ref{table:vlba_data}). 
These VLBA observations only cover about 8\% of the binary orbit and show no substantial increase in the flux density nor detectable outbursts.
Fig.~\ref{fig:flux_radio} shows that the closest radio observation to a periastron passage was that of VLASS2.1 at a frequency of 3\,GHz. This observation was obtained approximately one month after the periastron passage in 2020. It is unclear whether that observation shows flux perturbations, such as strong bursts.
Future multiwavelength observations near the time of periastron could show, for example, perturbations in the dips of the optical light curves and an increase in the flux density and the number and intensity of radio bursts.

\subsection{System distance and binary mass}

It is important to emphasize that the combined fit results are limited since the VLBA observations only cover a small fraction of the orbital period ($\sim$8\%) of both stars around their center of mass, and we used only four relative astrometric epochs and a fixed parallax. In addition, these observations do not provide enough information to estimate the masses of the individual components in this binary system. Furthermore, the parallax value provided in {\em Gaia} DR3 (20.703 mas) might be affected by the binary nature of the system, since {\em Gaia}'s observations do not resolve this compact binary system and do not properly cover its orbit. 
\citet[][]{miret20} proposed that 2M0508--21AB belongs to the $\beta$-Pictoris moving group, whose members span a wide range of distances between 10 and 70 pc \citep[e.g.,][]{binks16}. 
If 2M0508--21AB were located at a distance of 37.5\,pc---the median distance of the confirmed members of this moving group---the total mass and physical separation would be smaller by a factor of $\sim$1.7 ($M_{\rm tot}$ = 0.27\,M$_\sun$ versus 0.46\,M$_\sun$) and $\sim$1.2 ($a =$ 1.11\,au versus 1.30\,au), respectively, which are more consistent with theoretical evolutionary models \citep[e.g.,][]{baraffe15}. 
Therefore, it is crucial to obtain additional observations with the VLBA to achieve a reliable estimate of the parallax of the system and, consequently, the total mass of the binary system and the mass of the individual stars.

\section{Conclusions}\label{sec:conclusions}

The radio emission observed on milliarcsecond scales toward 2M0508--21AB is from both components, which were detected with flux densities between 0.3 and 0.7 mJy. Both stars have a similar flux density and exhibit similar flux-density variation in the three observed epochs. The similarity in the detected radio fluxes suggests that the two stars may be similar. This similarity is also observed in the values of their photometric periods (6.73 and 7.29 h, respectively) and their brightnesses in a Keck NIRC image obtained in the $H$-band. 

The radio maps at 4.85 GHz do not exhibit circular polarization (polarization fraction $\lesssim$10\%); however, the radio light curves of both stars show some tentative weak polarization only at two epochs and coinciding with maximum flux density. 
These results contrast with the relatively high circular polarization and circular polarized bursts seen through sub-gigahertz observations \citep{kaur24}. This suggests a general trend toward a lower polarization at higher frequencies. 

The combination of VLBA, Keck, and CARMENES data has proven highly effective in this project. The VLBA data provide excellent absolute astrometry of both stellar components, which we combined to obtain their relative astrometry. The Keck data provide less precise relative astrometry of the binary system but give an extended timeline (13 years) to anchor the fitted model. 
In addition, the CARMENES spectroscopic data provide the RV motion of the binary system over an eight-year baseline, covering more than three orbital periods of the system.
In combination, the three datasets provide the basis for a complete relative astrometric fit to an M-dwarf binary system with extraordinary precision. 
The combined fit shows that the orbit of the binary system is eccentric ($e = 0.714 \pm 0.0026$), with an orbital period of 2.194$\pm$0.003 yr and a semimajor axis of 1.303$\pm$0.012 au.
Although these datasets do not provide the information needed to obtain the dynamical mass of the individual stars, they allow for the determination of the dynamical mass of the binary system (M$_{\rm AB}$ $=$ 0.459$\pm$0.007 M$_{\odot}$), assuming that the distance derived from \textit{Gaia}'s parallax is correct.
Future VLBA observations, covering most of the binary orbital period, will help to confirm whether such a value of the parallax and the inferred total mass are correct or not and allow us to obtain the dynamical mass of the individual stars.

\begin{acknowledgements}
The authors thank the anonymous referee for providing very useful comments that improved this paper.
We acknowledge financial support from Universidad Nacional Aut\'onoma de M\'exico under grant UNAM-PAPIIT IN107324, 
from the Mexican Secretar\'ia de Ciencia, Humanidades, Tecnolog{\'\i}a e Innovaci\'on (Secihti) under grants CF-2023-I-232 and CBF-2025-I-201,
from the Agencia Estatal de Investigaci\'on (AEI/10.13039/501100011033) of the Spanish Ministerio de Ciencia, Innovaci\'on y Universidades (MICIU) and the European Regional Development Fund (ERDF) ``A way of making Europe'' through projects 
  PID2023-152906NA-I00,       
  PID2023-147883NB-C21,       
  PID2023-146675NB-I00,       
  PID2022-137241NB-C4[1:4],     
  PID2021-125627OB-C3[1:2],     
  RYC2021-032892-I,        
the Center of Excellence ``Severo Ochoa'' and ``Mar\'ia de Maeztu'' awards to the Instituto de Astrof\'isica de Andaluc\'ia (CEX2021-001131-S) and Institut de Ci\`encies de l'Espai (CEX2020-001058-M), the MaX-CSIC Excellence Award MaX4-SOMMA-ICE,
and AST22\_00001\_Subp~8 (co-funded by the European Union via NextGenerationEU, the Spanish Consejo Superior de Investigaciones Cient\'ificas, and the Agencia de Innovaci\'on y Desarrollo de Andaluc\'ia),
from the European Research Council (ERC) under the European Union’s Horizon 2020 research and innovation program (ERC-StG ``IMAGINE'' No. 948582 and ERC-StG ``STORM-CHASER'' No. 101042416),
and from the European Cooperation in Science and Technology (COST Action ``PLANETS'' No. CA22133). 

Part of this work was carried out within the framework of the doctoral program in Physics of the Universitat Aut\`onoma de Barcelona. 
 
The observations were carried out with the Very Long Baseline Array (VLBA), which is part of the National Radio Astronomy Observatory (NRAO). The NRAO is a facility of the National Science Foundation, operated under a cooperative agreement by Associated Universities, Inc. 
This publication used the SIMBAD database operated at the Centre de donn\'ees astronomiques de Strasbourg, France. 

Software: {\tt AIPS}  \citep[][]{greisen03}, {\tt astropy} \citep[][]{astropy13,astropy18}, {\tt corner} \citep[][]{foremanmackey16}, {\tt emcee} \citep[][]{foremanmackey13}, {\tt lmfit} \citep[][]{newville20}, {\tt scipy}  \citep[][]{2020SciPy-NMeth}, {\tt matplotlib} \citep[][]{hunter07}, and {\tt numpy}  \citep[][]{vanderwalt11}.

\end{acknowledgements}

\bibliographystyle{aa} 
\bibliography{radio}

@ARTICLE{astropy13,
       author = {{Astropy Collaboration} and {Robitaille}, Thomas P. and {Tollerud}, Erik J. and {Greenfield}, Perry and {Droettboom}, Michael and {Bray}, Erik and {Aldcroft}, Tom and {Davis}, Matt and {Ginsburg}, Adam and {Price-Whelan}, Adrian M. and {Kerzendorf}, Wolfgang E. and {Conley}, Alexander and {Crighton}, Neil and {Barbary}, Kyle and {Muna}, Demitri and {Ferguson}, Henry and {Grollier}, Fr{\'e}d{\'e}ric and {Parikh}, Madhura M. and {Nair}, Prasanth H. and {Unther}, Hans M. and {Deil}, Christoph and {Woillez}, Julien and {Conseil}, Simon and {Kramer}, Roban and {Turner}, James E.~H. and {Singer}, Leo and {Fox}, Ryan and {Weaver}, Benjamin A. and {Zabalza}, Victor and {Edwards}, Zachary I. and {Azalee Bostroem}, K. and {Burke}, D.~J. and {Casey}, Andrew R. and {Crawford}, Steven M. and {Dencheva}, Nadia and {Ely}, Justin and {Jenness}, Tim and {Labrie}, Kathleen and {Lim}, Pey Lian and {Pierfederici}, Francesco and {Pontzen}, Andrew and {Ptak}, Andy and {Refsdal}, Brian and {Servillat}, Mathieu and {Streicher}, Ole},
        title = "{Astropy: A community Python package for astronomy}",
      journal = {\aap},
         year = 2013,
        month = oct,
       volume = {558},
          eid = {A33},
        pages = {A33},
          doi = {10.1051/0004-6361/201322068},
archivePrefix = {arXiv},
       eprint = {1307.6212},
 primaryClass = {astro-ph.IM},
       adsurl = {https://ui.adsabs.harvard.edu/abs/2013A&A...558A..33A}
}

@ARTICLE{astropy18,
       author = {{Astropy Collaboration} and {Price-Whelan}, A.~M. and {Sip{\H{o}}cz}, B.~M. and {G{\"u}nther}, H.~M. and {Lim}, P.~L. and {Crawford}, S.~M. and {Conseil}, S. and {Shupe}, D.~L. and {Craig}, M.~W. and {Dencheva}, N. and {Ginsburg}, A. and {VanderPlas}, J.~T. and {Bradley}, L.~D. and {P{\'e}rez-Su{\'a}rez}, D. and {de Val-Borro}, M. and {Aldcroft}, T.~L. and {Cruz}, K.~L. and {Robitaille}, T.~P. and {Tollerud}, E.~J. and {Ardelean}, C. and {Babej}, T. and {Bach}, Y.~P. and {Bachetti}, M. and {Bakanov}, A.~V. and {Bamford}, S.~P. and {Barentsen}, G. and {Barmby}, P. and {Baumbach}, A. and {Berry}, K.~L. and {Biscani}, F. and {Boquien}, M. and {Bostroem}, K.~A. and {Bouma}, L.~G. and {Brammer}, G.~B. and {Bray}, E.~M. and {Breytenbach}, H. and {Buddelmeijer}, H. and {Burke}, D.~J. and {Calderone}, G. and {Cano Rodr{\'\i}guez}, J.~L. and {Cara}, M. and {Cardoso}, J.~V.~M. and {Cheedella}, S. and {Copin}, Y. and {Corrales}, L. and {Crichton}, D. and {D'Avella}, D. and {Deil}, C. and {Depagne}, {\'E}. and {Dietrich}, J.~P. and {Donath}, A. and {Droettboom}, M. and {Earl}, N. and {Erben}, T. and {Fabbro}, S. and {Ferreira}, L.~A. and {Finethy}, T. and {Fox}, R.~T. and {Garrison}, L.~H. and {Gibbons}, S.~L.~J. and {Goldstein}, D.~A. and {Gommers}, R. and {Greco}, J.~P. and {Greenfield}, P. and {Groener}, A.~M. and {Grollier}, F. and {Hagen}, A. and {Hirst}, P. and {Homeier}, D. and {Horton}, A.~J. and {Hosseinzadeh}, G. and {Hu}, L. and {Hunkeler}, J.~S. and {Ivezi{\'c}}, {\v{Z}}. and {Jain}, A. and {Jenness}, T. and {Kanarek}, G. and {Kendrew}, S. and {Kern}, N.~S. and {Kerzendorf}, W.~E. and {Khvalko}, A. and {King}, J. and {Kirkby}, D. and {Kulkarni}, A.~M. and {Kumar}, A. and {Lee}, A. and {Lenz}, D. and {Littlefair}, S.~P. and {Ma}, Z. and {Macleod}, D.~M. and {Mastropietro}, M. and {McCully}, C. and {Montagnac}, S. and {Morris}, B.~M. and {Mueller}, M. and {Mumford}, S.~J. and {Muna}, D. and {Murphy}, N.~A. and {Nelson}, S. and {Nguyen}, G.~H. and {Ninan}, J.~P. and {N{\"o}the}, M. and {Ogaz}, S. and {Oh}, S. and {Parejko}, J.~K. and {Parley}, N. and {Pascual}, S. and {Patil}, R. and {Patil}, A.~A. and {Plunkett}, A.~L. and {Prochaska}, J.~X. and {Rastogi}, T. and {Reddy Janga}, V. and {Sabater}, J. and {Sakurikar}, P. and {Seifert}, M. and {Sherbert}, L.~E. and {Sherwood-Taylor}, H. and {Shih}, A.~Y. and {Sick}, J. and {Silbiger}, M.~T. and {Singanamalla}, S. and {Singer}, L.~P. and {Sladen}, P.~H. and {Sooley}, K.~A. and {Sornarajah}, S. and {Streicher}, O. and {Teuben}, P. and {Thomas}, S.~W. and {Tremblay}, G.~R. and {Turner}, J.~E.~H. and {Terr{\'o}n}, V. and {van Kerkwijk}, M.~H. and {de la Vega}, A. and {Watkins}, L.~L. and {Weaver}, B.~A. and {Whitmore}, J.~B. and {Woillez}, J. and {Zabalza}, V. and {Astropy Contributors}},
        title = "{The Astropy Project: Building an Open-science Project and Status of the v2.0 Core Package}",
      journal = {\aj},
         year = 2018,
        month = sep,
       volume = {156},
       number = {3},
          eid = {123},
        pages = {123},
          doi = {10.3847/1538-3881/aabc4f},
archivePrefix = {arXiv},
       eprint = {1801.02634},
 primaryClass = {astro-ph.IM},
       adsurl = {https://ui.adsabs.harvard.edu/abs/2018AJ....156..123A}
}

@ARTICLE{babusiaux23,
       author = {{Babusiaux}, C. and {Fabricius}, C. and {Khanna}, S. and {Muraveva}, T. and {Reyl{\'e}}, C. and {Spoto}, F. and {Vallenari}, A. and {Luri}, X. and {Arenou}, F. and {{\'A}lvarez}, M.~A. and {Anders}, F. and {Antoja}, T. and {Balbinot}, E. and {Barache}, C. and {Bauchet}, N. and {Bossini}, D. and {Busonero}, D. and {Cantat-Gaudin}, T. and {Carrasco}, J.~M. and {Dafonte}, C. and {Diakit{\'e}}, S. and {Figueras}, F. and {Garcia-Gutierrez}, A. and {Garofalo}, A. and {Helmi}, A. and {Jim{\'e}nez-Arranz}, {\'O}. and {Jordi}, C. and {Kervella}, P. and {Kostrzewa-Rutkowska}, Z. and {Leclerc}, N. and {Licata}, E. and {Manteiga}, M. and {Masip}, A. and {Mongui{\'o}}, M. and {Ramos}, P. and {Robichon}, N. and {Robin}, A.~C. and {Romero-G{\'o}mez}, M. and {S{\'a}ez}, A. and {Santove{\~n}a}, R. and {Spina}, L. and {Torralba Elipe}, G. and {Weiler}, M.},
        title = "{Gaia Data Release 3. Catalogue validation}",
      journal = {\aap},
         year = 2023,
        month = jun,
       volume = {674},
          eid = {A32},
        pages = {A32},
          doi = {10.1051/0004-6361/202243790},
archivePrefix = {arXiv},
       eprint = {2206.05989},
 primaryClass = {astro-ph.SR},
       adsurl = {https://ui.adsabs.harvard.edu/abs/2023A&A...674A..32B}
}

@ARTICLE{baraffe15,
       author = {{Baraffe}, Isabelle and {Homeier}, Derek and {Allard}, France and {Chabrier}, Gilles},
        title = "{New evolutionary models for pre-main sequence and main sequence low-mass stars down to the hydrogen-burning limit}",
      journal = {\aap},
         year = 2015,
        month = may,
       volume = {577},
          eid = {A42},
        pages = {A42},
          doi = {10.1051/0004-6361/201425481},
archivePrefix = {arXiv},
       eprint = {1503.04107},
 primaryClass = {astro-ph.SR},
       adsurl = {https://ui.adsabs.harvard.edu/abs/2015A&A...577A..42B}
}

@ARTICLE{berger01,
       author = {{Berger}, E. and {Ball}, S. and {Becker}, K.~M. and {Clarke}, M. and {Frail}, D.~A. and {Fukuda}, T.~A. and {Hoffman}, I.~M. and {Mellon}, R. and {Momjian}, E. and {Murphy}, N.~W. and {Teng}, S.~H. and {Woodruff}, T. and {Zauderer}, B.~A. and {Zavala}, R.~T.},
        title = "{Discovery of radio emission from the brown dwarf LP944-20}",
      journal = {\nat},
         year = 2001,
        month = mar,
       volume = {410},
       number = {6826},
        pages = {338-340},
archivePrefix = {arXiv},
       eprint = {astro-ph/0102301},
 primaryClass = {astro-ph},
       adsurl = {https://ui.adsabs.harvard.edu/abs/2001Natur.410..338B}
}

@ARTICLE{berger02,
       author = {{Berger}, E.},
        title = "{Flaring up All Over-Radio Activity in Rapidly Rotating Late M and L Dwarfs}",
      journal = {\apj},
         year = 2002,
        month = jun,
       volume = {572},
       number = {1},
        pages = {503-513},
          doi = {10.1086/340301},
archivePrefix = {arXiv},
       eprint = {astro-ph/0111317},
 primaryClass = {astro-ph},
       adsurl = {https://ui.adsabs.harvard.edu/abs/2002ApJ...572..503B}
}

@ARTICLE{berger06,
       author = {{Berger}, E.},
        title = "{Radio Observations of a Large Sample of Late M, L, and T Dwarfs: The Distribution of Magnetic Field Strengths}",
      journal = {\apj},
         year = 2006,
        month = sep,
       volume = {648},
       number = {1},
        pages = {629-636},
          doi = {10.1086/505787},
archivePrefix = {arXiv},
       eprint = {astro-ph/0603176},
 primaryClass = {astro-ph},
       adsurl = {https://ui.adsabs.harvard.edu/abs/2006ApJ...648..629B}
}

@ARTICLE{binks16,
       author = {{Binks}, A.~S. and {Jeffries}, R.~D.},
        title = "{Spectroscopic confirmation of M-dwarf candidate members of the Beta Pictoris and AB Doradus Moving Groups}",
      journal = {\mnras},
         year = 2016,
        month = jan,
       volume = {455},
       number = {3},
        pages = {3345-3358},
          doi = {10.1093/mnras/stv2431},
archivePrefix = {arXiv},
       eprint = {1510.06987},
 primaryClass = {astro-ph.SR},
       adsurl = {https://ui.adsabs.harvard.edu/abs/2016MNRAS.455.3345B}
}

@article{bloot24,
       author = {{Bloot}, S. and {Callingham}, J.~R. and {Vedantham}, H.~K. and {Kavanagh}, R.~D. and {Pope}, B.~J.~S. and {Climent}, J.~B. and {Guirado}, J.~C. and {Pe{\~n}a-Mo{\~n}ino}, L. and {P{\'e}rez-Torres}, M.},
        title = "{Phenomenology and periodicity of radio emission from the stellar system AU Microscopii}",
      journal = {\aap},
         year = 2024,
        month = feb,
       volume = {682},
          eid = {A170},
        pages = {A170},
          doi = {10.1051/0004-6361/202348065},
archivePrefix = {arXiv},
       eprint = {2312.09071},
 primaryClass = {astro-ph.SR},
       adsurl = {https://ui.adsabs.harvard.edu/abs/2024A&A...682A.170B}
}

@ARTICLE{bouma24,
       author = {{Bouma}, Luke G. and {Jayaraman}, Rahul and {Rappaport}, Saul and {Rebull}, Luisa M. and {Hillenbrand}, Lynne A. and {Winn}, Joshua N. and {David-Uraz}, Alexandre and {Bakos}, G{\'a}sp{\'a}r {\'A}.},
        title = "{Transient Corotating Clumps around Adolescent Low-mass Stars from Four Years of TESS}",
      journal = {\aj},
         year = 2024,
        month = jan,
       volume = {167},
       number = {1},
          eid = {38},
        pages = {38},
          doi = {10.3847/1538-3881/ad0c4c},
archivePrefix = {arXiv},
       eprint = {2309.06471},
 primaryClass = {astro-ph.SR},
       adsurl = {https://ui.adsabs.harvard.edu/abs/2024AJ....167...38B}
}

@ARTICLE{bouma25,
       author = {{Bouma}, Luke G. and {Jardine}, Moira M.},
        title = "{A Plasma Torus around a Young Low-mass Star}",
      journal = {\apjl},
         year = 2025,
        month = jul,
       volume = {988},
       number = {1},
          eid = {L3},
        pages = {L3},
          doi = {10.3847/2041-8213/ade39a},
archivePrefix = {arXiv},
       eprint = {2506.09116},
 primaryClass = {astro-ph.SR},
       adsurl = {https://ui.adsabs.harvard.edu/abs/2025ApJ...988L...3B}
}

@ARTICLE{burgasser15,
       author = {{Burgasser}, Adam J. and {Melis}, Carl and {Todd}, Jacob and {Gelino}, Christopher R. and {Hallinan}, Gregg and {Bardalez Gagliuffi}, Daniella},
        title = "{Radio Emission and Orbital Motion from the Close-encounter Star-Brown Dwarf Binary WISE J072003.20-084651.2}",
      journal = {\aj},
         year = 2015,
        month = dec,
       volume = {150},
       number = {6},
          eid = {180},
        pages = {180},
          doi = {10.1088/0004-6256/150/6/180},
archivePrefix = {arXiv},
       eprint = {1508.06332},
 primaryClass = {astro-ph.SR},
       adsurl = {https://ui.adsabs.harvard.edu/abs/2015AJ....150..180B}
}

@INPROCEEDINGS{caballero16,
       author = {{Caballero}, J.~A. and {Gu{\`a}rdia}, J. and {L{\'o}pez del Fresno}, M. and {Zechmeister}, M. and {de Juan}, E. and {Alonso-Floriano}, F.~J. and {Amado}, P.~J. and {Colom{\'e}}, J. and {Cort{\'e}s-Contreras}, M. and {Garc{\'\i}a-Piquer}, {\'A}. and {Gesa}, L. and {de Guindos}, E. and {Hagen}, H. -J. and {Helmling}, J. and {Hern{\'a}ndez Casta{\~n}o}, L. and {K{\"u}rster}, M. and {L{\'o}pez-Santiago}, J. and {Montes}, D. and {Morales Mu{\~n}oz}, R. and {Pavlov}, A. and {Quirrenbach}, A. and {Reiners}, A. and {Ribas}, I. and {Seifert}, W. and {Solano}, E.},
        title = "{CARMENES: data flow}",
    booktitle = {Observatory Operations: Strategies, Processes, and Systems VI},
         year = 2016,
       editor = {{Peck}, Alison B. and {Seaman}, Robert L. and {Benn}, Chris R.},
       series = {Society of Photo-Optical Instrumentation Engineers (SPIE) Conference Series},
       volume = {9910},
        month = jul,
          eid = {99100E},
        pages = {99100E},
          doi = {10.1117/12.2233574},
       adsurl = {https://ui.adsabs.harvard.edu/abs/2016SPIE.9910E..0EC}
}

@ARTICLE{calissendorff22,
       author = {{Calissendorff}, Per and {Janson}, Markus and {Rodet}, Laetitia and {K{\"o}hler}, Rainer and {Bonnefoy}, Micka{\"e}l and {Brandner}, Wolfgang and {Brown-Sevilla}, Samantha and {Chauvin}, Ga{\"e}l and {Delorme}, Philippe and {Desidera}, Silvano and {Durkan}, Stephen and {Fontanive}, Clemence and {Gratton}, Raffaele and {Hagelberg}, Janis and {Henning}, Thomas and {Hippler}, Stefan and {Lagrange}, Anne-Marie and {Langlois}, Maud and {Lazzoni}, Cecilia and {Maire}, Anne-Lise and {Messina}, Sergio and {Meyer}, Michael and {M{\"o}ller-Nilsson}, Ole and {Rabus}, Markus and {Schlieder}, Joshua and {Vigan}, Arthur and {Wahhaj}, Zahed and {Wildi}, Francois and {Zurlo}, Alice},
        title = "{Updated orbital monitoring and dynamical masses for nearby M-dwarf binaries}",
      journal = {\aap},
         year = 2022,
        month = oct,
       volume = {666},
          eid = {A16},
        pages = {A16},
          doi = {10.1051/0004-6361/202142766},
archivePrefix = {arXiv},
       eprint = {2208.09503},
 primaryClass = {astro-ph.SR},
       adsurl = {https://ui.adsabs.harvard.edu/abs/2022A&A...666A..16C}
}

@ARTICLE{callingham21,
       author = {{Callingham}, J.~R. and {Vedantham}, H.~K. and {Shimwell}, T.~W. and {Pope}, B.~J.~S. and {Davis}, I.~E. and {Best}, P.~N. and {Hardcastle}, M.~J. and {R{\"o}ttgering}, H.~J.~A. and {Sabater}, J. and {Tasse}, C. and {van Weeren}, R.~J. and {Williams}, W.~L. and {Zarka}, P. and {de Gasperin}, F. and {Drabent}, A.},
        title = "{The population of M dwarfs observed at low radio frequencies}",
      journal = {Nat.As},
         year = 2021,
        month = oct,
       volume = {5},
        pages = {1233-1239},
          doi = {10.1038/s41550-021-01483-0},
archivePrefix = {arXiv},
       eprint = {2110.03713},
 primaryClass = {astro-ph.SR},
       adsurl = {https://ui.adsabs.harvard.edu/abs/2021NatAs...5.1233C}
}

@ARTICLE{callingham24,
       author = {{Callingham}, J.~R. and {Pope}, B.~J.~S. and {Kavanagh}, R.~D. and {Bellotti}, S. and {Daley-Yates}, S. and {Damasso}, M. and {Grie{\ss}meier}, J. -M. and {G{\"u}del}, M. and {G{\"u}nther}, M. and {Kao}, M.~M. and {Klein}, B. and {Mahadevan}, S. and {Morin}, J. and {Nichols}, J.~D. and {Osten}, R.~A. and {P{\'e}rez-Torres}, M. and {Pineda}, J.~S. and {Rigney}, J. and {Saur}, J. and {Stef{\'a}nsson}, G. and {Turner}, J.~D. and {Vedantham}, H. and {Vidotto}, A.~A. and {Villadsen}, J. and {Zarka}, P.},
        title = "{Radio signatures of star-planet interactions, exoplanets and space weather}",
      journal = {Nature Astronomy},
         year = 2024,
        month = nov,
       volume = {8},
        pages = {1359-1372},
          doi = {10.1038/s41550-024-02405-6},
archivePrefix = {arXiv},
       eprint = {2409.15507},
 primaryClass = {astro-ph.EP},
       adsurl = {https://ui.adsabs.harvard.edu/abs/2024NatAs...8.1359C}
}

@ARTICLE{callingham25,
       author = {{Callingham}, J.~R. and {Tasse}, C. and {Keers}, R. and {Kavanagh}, R.~D. and {Vedantham}, H.~K. and {Zarka}, P. and {Bellotti}, S. and {Cristofari}, P.~I. and {Bloot}, S. and {Konijn}, D.~C. and {Hardcastle}, M.~J. and {Lamy}, L. and {Pass}, E.~K. and {Pope}, B.~J.~S. and {Reid}, H. and {R{\"o}ttgering}, H.~J.~A. and {Shimwell}, T.~W. and {Zucca}, P.},
        title = "{Radio burst from a stellar coronal mass ejection}",
      journal = {\nat},
         year = 2025,
        month = nov,
       volume = {647},
       number = {8090},
        pages = {603-607},
          doi = {10.1038/s41586-025-09715-3},
archivePrefix = {arXiv},
       eprint = {2511.09289},
 primaryClass = {astro-ph.SR},
       adsurl = {https://ui.adsabs.harvard.edu/abs/2025Natur.647..603C}
}

@ARTICLE{2020SciPy-NMeth,
  author  = {Virtanen, Pauli and Gommers, Ralf and Oliphant, Travis E. and
            Haberland, Matt and Reddy, Tyler and Cournapeau, David and
            Burovski, Evgeni and Peterson, Pearu and Weckesser, Warren and
            Bright, Jonathan and {van der Walt}, St{\'e}fan J. and
            Brett, Matthew and Wilson, Joshua and Millman, K. Jarrod and
            Mayorov, Nikolay and Nelson, Andrew R. J. and Jones, Eric and
            Kern, Robert and Larson, Eric and Carey, C J and
            Polat, {\.I}lhan and Feng, Yu and Moore, Eric W. and
            {VanderPlas}, Jake and Laxalde, Denis and Perktold, Josef and
            Cimrman, Robert and Henriksen, Ian and Quintero, E. A. and
            Harris, Charles R. and Archibald, Anne M. and
            Ribeiro, Ant{\^o}nio H. and Pedregosa, Fabian and
            {van Mulbregt}, Paul and {SciPy 1.0 Contributors}},
  title   = {{{SciPy} 1.0: Fundamental Algorithms for Scientific
            Computing in Python}},
  journal = {Nature Methods},
  year    = {2020},
  volume  = {17},
  pages   = {261--272},
  adsurl  = {https://rdcu.be/b08Wh},
  doi     = {10.1038/s41592-019-0686-2},
}

@ARTICLE{clark80,
       author = {{Clark}, B.~G.},
        title = "{An efficient implementation of the algorithm 'CLEAN'}",
      journal = {\aap},
         year = 1980,
        month = sep,
       volume = {89},
       number = {3},
        pages = {377},
       adsurl = {https://ui.adsabs.harvard.edu/abs/1980A&A....89..377C}
}

@ARTICLE{climent23,
       author = {{Climent}, J.~B. and {Guirado}, J.~C. and {P{\'e}rez-Torres}, M. and {Marcaide}, J.~M. and {Pe{\~n}a-Mo{\~n}ino}, L.},
        title = "{Evidence for a radiation belt around a brown dwarf}",
      journal = {Science},
         year = 2023,
        month = sep,
       volume = {381},
       number = {6662},
        pages = {1120-1124},
          doi = {10.1126/science.adg6635},
archivePrefix = {arXiv},
       eprint = {2303.06453},
 primaryClass = {astro-ph.SR},
       adsurl = {https://ui.adsabs.harvard.edu/abs/2023Sci...381.1120C}
}

@ARTICLE{curiel20,
       author = {{Curiel}, Salvador and {Ortiz-Le{\'o}n}, Gisela N. and {Mioduszewski}, Amy J. and {Torres}, Rosa M.},
        title = "{An Astrometric Planetary Companion Candidate to the M9 Dwarf TVLM 513-46546}",
      journal = {\aj},
         year = 2020,
        month = sep,
       volume = {160},
       number = {3},
          eid = {97},
        pages = {97},
          doi = {10.3847/1538-3881/ab9e6e},
archivePrefix = {arXiv},
       eprint = {2008.01595},
 primaryClass = {astro-ph.EP},
       adsurl = {https://ui.adsabs.harvard.edu/abs/2020AJ....160...97C}
}

@ARTICLE{curiel22,
       author = {{Curiel}, Salvador and {Ortiz-Le{\'o}n}, Gisela N. and {Mioduszewski}, Amy J. and {Sanchez-Bermudez}, Joel},
        title = "{3D Orbital Architecture of a Dwarf Binary System and Its Planetary Companion}",
      journal = {\aj},
         year = 2022,
        month = sep,
       volume = {164},
       number = {3},
          eid = {93},
        pages = {93},
          doi = {10.3847/1538-3881/ac7c66},
archivePrefix = {arXiv},
       eprint = {2208.14553},
 primaryClass = {astro-ph.EP},
       adsurl = {https://ui.adsabs.harvard.edu/abs/2022AJ....164...93C}
}

@ARTICLE{curiel24,
       author = {{Curiel}, Salvador and {Ortiz-Le{\'o}n}, Gisela N. and {Mioduszewski}, Amy J. and {Arenas-Martinez}, Anthony B.},
        title = "{Precise Mass, Orbital Motion, and Stellar Properties of the M-dwarf Binary LP 349‑25AB}",
      journal = {\apj},
         year = 2024,
        month = jun,
       volume = {967},
       number = {2},
          eid = {112},
        pages = {112},
          doi = {10.3847/1538-4357/ad3df6},
archivePrefix = {arXiv},
       eprint = {2404.16964},
 primaryClass = {astro-ph.SR},
       adsurl = {https://ui.adsabs.harvard.edu/abs/2024ApJ...967..112C}
}

@ARTICLE{driessen24,
       author = {{Driessen}, Laura Nicole and {Pritchard}, Joshua and {Murphy}, Tara and {Heald}, George and {Robrade}, Jan and {Das}, Barnali and {Duchesne}, Stefan William and {Kaplan}, David L. and {Lenc}, Emil and {Lynch}, Christene R. and {Mitchell-Bolton}, Jackson and {Pope}, Benjamin J.~S. and {Rose}, Kovi and {Stelzer}, Beate and {Wang}, Yuanming and {Zic}, Andrew},
        title = "{The Sydney Radio Star Catalogue: Properties of radio stars at megahertz to gigahertz frequencies}",
      journal = {\pasa},
         year = 2024,
        month = nov,
       volume = {41},
          eid = {e084},
        pages = {e084},
          doi = {10.1017/pasa.2024.72},
archivePrefix = {arXiv},
       eprint = {2404.07418},
 primaryClass = {astro-ph.SR},
       adsurl = {https://ui.adsabs.harvard.edu/abs/2024PASA...41...84D}
}

@ARTICLE{dulk85,
       author = {{Dulk}, G.~A.},
        title = "{Radio emission from the sun and stars.}",
      journal = {\araa},
         year = 1985,
        month = jan,
       volume = {23},
        pages = {169-224},
          doi = {10.1146/annurev.aa.23.090185.001125},
       adsurl = {https://ui.adsabs.harvard.edu/abs/1985ARA&A..23..169D}
}

@ARTICLE{dupuy16,
       author = {{Dupuy}, Trent J. and {Forbrich}, Jan and {Rizzuto}, Aaron and {Mann}, Andrew W. and {Aller}, Kimberly and {Liu}, Michael C. and {Kraus}, Adam L. and {Berger}, Edo},
        title = "{High-precision Radio and Infrared Astrometry of LSPM J1314+1320AB. II. Testing Pre-main-sequence Models at the Lithium Depletion Boundary with Dynamical Masses}",
      journal = {\apj},
         year = 2016,
        month = aug,
       volume = {827},
       number = {1},
          eid = {23},
        pages = {23},
          doi = {10.3847/0004-637X/827/1/23},
archivePrefix = {arXiv},
       eprint = {1605.07182},
 primaryClass = {astro-ph.SR},
       adsurl = {https://ui.adsabs.harvard.edu/abs/2016ApJ...827...23D}
}

@ARTICLE{foremanmackey13,
       author = {{Foreman-Mackey}, Daniel and {Hogg}, David W. and {Lang}, Dustin and {Goodman}, Jonathan},
        title = "{emcee: The MCMC Hammer}",
      journal = {\pasp},
         year = 2013,
        month = mar,
       volume = {125},
       number = {925},
        pages = {306},
          doi = {10.1086/670067},
archivePrefix = {arXiv},
       eprint = {1202.3665},
 primaryClass = {astro-ph.IM},
       adsurl = {https://ui.adsabs.harvard.edu/abs/2013PASP..125..306F}
}

@ARTICLE{foremanmackey16,
       author = {{Foreman-Mackey}, Daniel},
        title = "{corner.py: Scatterplot matrices in Python}",
      journal = {The Journal of Open Source Software},
         year = 2016,
        month = jun,
       volume = {1},
        pages = {24},
          doi = {10.21105/joss.00024},
       adsurl = {https://ui.adsabs.harvard.edu/abs/2016JOSS....1...24F}
}

@ARTICLE{gaia21,
       author = {{Gaia Collaboration} and {Brown}, A.~G.~A. and {Vallenari}, A. and {Prusti}, T. and {de Bruijne}, J.~H.~J. and {Babusiaux}, C. and {Biermann}, M. and {Creevey}, O.~L. and {Evans}, D.~W. and {Eyer}, L. and {Hutton}, A. and {Jansen}, F. and {Jordi}, C. and {Klioner}, S.~A. and {Lammers}, U. and {Lindegren}, L. and {Luri}, X. and {Mignard}, F. and {Panem}, C. and {Pourbaix}, D. and {Randich}, S. and {Sartoretti}, P. and {Soubiran}, C. and {Walton}, N.~A. and {Arenou}, F. and {Bailer-Jones}, C.~A.~L. and {Bastian}, U. and {Cropper}, M. and {Drimmel}, R. and {Katz}, D. and {Lattanzi}, M.~G. and {van Leeuwen}, F. and {Bakker}, J. and {Cacciari}, C. and {Casta{\~n}eda}, J. and {De Angeli}, F. and {Ducourant}, C. and {Fabricius}, C. and {Fouesneau}, M. and {Fr{\'e}mat}, Y. and {Guerra}, R. and {Guerrier}, A. and {Guiraud}, J. and {Jean-Antoine Piccolo}, A. and {Masana}, E. and {Messineo}, R. and {Mowlavi}, N. and {Nicolas}, C. and {Nienartowicz}, K. and {Pailler}, F. and {Panuzzo}, P. and {Riclet}, F. and {Roux}, W. and {Seabroke}, G.~M. and {Sordo}, R. and {Tanga}, P. and {Th{\'e}venin}, F. and {Gracia-Abril}, G. and {Portell}, J. and {Teyssier}, D. and {Altmann}, M. and {Andrae}, R. and {Bellas-Velidis}, I. and {Benson}, K. and {Berthier}, J. and {Blomme}, R. and {Brugaletta}, E. and {Burgess}, P.~W. and {Busso}, G. and {Carry}, B. and {Cellino}, A. and {Cheek}, N. and {Clementini}, G. and {Damerdji}, Y. and {Davidson}, M. and {Delchambre}, L. and {Dell'Oro}, A. and {Fern{\'a}ndez-Hern{\'a}ndez}, J. and {Galluccio}, L. and {Garc{\'\i}a-Lario}, P. and {Garcia-Reinaldos}, M. and {Gonz{\'a}lez-N{\'u}{\~n}ez}, J. and {Gosset}, E. and {Haigron}, R. and {Halbwachs}, J. -L. and {Hambly}, N.~C. and {Harrison}, D.~L. and {Hatzidimitriou}, D. and {Heiter}, U. and {Hern{\'a}ndez}, J. and {Hestroffer}, D. and {Hodgkin}, S.~T. and {Holl}, B. and {Jan{\ss}en}, K. and {Jevardat de Fombelle}, G. and {Jordan}, S. and {Krone-Martins}, A. and {Lanzafame}, A.~C. and {L{\"o}ffler}, W. and {Lorca}, A. and {Manteiga}, M. and {Marchal}, O. and {Marrese}, P.~M. and {Moitinho}, A. and {Mora}, A. and {Muinonen}, K. and {Osborne}, P. and {Pancino}, E. and {Pauwels}, T. and {Petit}, J. -M. and {Recio-Blanco}, A. and {Richards}, P.~J. and {Riello}, M. and {Rimoldini}, L. and {Robin}, A.~C. and {Roegiers}, T. and {Rybizki}, J. and {Sarro}, L.~M. and {Siopis}, C. and {Smith}, M. and {Sozzetti}, A. and {Ulla}, A. and {Utrilla}, E. and {van Leeuwen}, M. and {van Reeven}, W. and {Abbas}, U. and {Abreu Aramburu}, A. and {Accart}, S. and {Aerts}, C. and {Aguado}, J.~J. and {Ajaj}, M. and {Altavilla}, G. and {{\'A}lvarez}, M.~A. and {{\'A}lvarez Cid-Fuentes}, J. and {Alves}, J. and {Anderson}, R.~I. and {Anglada Varela}, E. and {Antoja}, T. and {Audard}, M. and {Baines}, D. and {Baker}, S.~G. and {Balaguer-N{\'u}{\~n}ez}, L. and {Balbinot}, E. and {Balog}, Z. and {Barache}, C. and {Barbato}, D. and {Barros}, M. and {Barstow}, M.~A. and {Bartolom{\'e}}, S. and {Bassilana}, J. -L. and {Bauchet}, N. and {Baudesson-Stella}, A. and {Becciani}, U. and {Bellazzini}, M. and {Bernet}, M. and {Bertone}, S. and {Bianchi}, L. and {Blanco-Cuaresma}, S. and {Boch}, T. and {Bombrun}, A. and {Bossini}, D. and {Bouquillon}, S. and {Bragaglia}, A. and {Bramante}, L. and {Breedt}, E. and {Bressan}, A. and {Brouillet}, N. and {Bucciarelli}, B. and {Burlacu}, A. and {Busonero}, D. and {Butkevich}, A.~G. and {Buzzi}, R. and {Caffau}, E. and {Cancelliere}, R. and {C{\'a}novas}, H. and {Cantat-Gaudin}, T. and {Carballo}, R. and {Carlucci}, T. and {Carnerero}, M.~I. and {Carrasco}, J.~M. and {Casamiquela}, L. and {Castellani}, M. and {Castro-Ginard}, A. and {Castro Sampol}, P. and {Chaoul}, L. and {Charlot}, P. and {Chemin}, L. and {Chiavassa}, A. and {Cioni}, M. -R.~L. and {Comoretto}, G. and {Cooper}, W.~J. and {Cornez}, T. and {Cowell}, S. and {Crifo}, F. and {Crosta}, M. and {Crowley}, C. and {Dafonte}, C. and {Dapergolas}, A. and {David}, M. and {David}, P. and {de Laverny}, P. and {De Luise}, F. and {De March}, R. and {De Ridder}, J. and {de Souza}, R. and {de Teodoro}, P. and {de Torres}, A. and {del Peloso}, E.~F. and {del Pozo}, E. and {Delbo}, M. and {Delgado}, A. and {Delgado}, H.~E. and {Delisle}, J. -B. and {Di Matteo}, P. and {Diakite}, S. and {Diener}, C. and {Distefano}, E. and {Dolding}, C. and {Eappachen}, D. and {Edvardsson}, B. and {Enke}, H. and {Esquej}, P. and {Fabre}, C. and {Fabrizio}, M. and {Faigler}, S. and {Fedorets}, G. and {Fernique}, P. and {Fienga}, A. and {Figueras}, F. and {Fouron}, C. and {Fragkoudi}, F. and {Fraile}, E. and {Franke}, F. and {Gai}, M. and {Garabato}, D. and {Garcia-Gutierrez}, A. and {Garc{\'\i}a-Torres}, M. and {Garofalo}, A. and {Gavras}, P. and {Gerlach}, E. and {Geyer}, R. and {Giacobbe}, P. and {Gilmore}, G. and {Girona}, S. and {Giuffrida}, G. and {Gomel}, R. and {Gomez}, A. and {Gonzalez-Santamaria}, I. and {Gonz{\'a}lez-Vidal}, J.~J. and {Granvik}, M. and {Guti{\'e}rrez-S{\'a}nchez}, R. and {Guy}, L.~P. and {Hauser}, M. and {Haywood}, M. and {Helmi}, A. and {Hidalgo}, S.~L. and {Hilger}, T. and {H{\l}adczuk}, N. and {Hobbs}, D. and {Holland}, G. and {Huckle}, H.~E. and {Jasniewicz}, G. and {Jonker}, P.~G. and {Juaristi Campillo}, J. and {Julbe}, F. and {Karbevska}, L. and {Kervella}, P. and {Khanna}, S. and {Kochoska}, A. and {Kontizas}, M. and {Kordopatis}, G. and {Korn}, A.~J. and {Kostrzewa-Rutkowska}, Z. and {Kruszy{\'n}ska}, K. and {Lambert}, S. and {Lanza}, A.~F. and {Lasne}, Y. and {Le Campion}, J. -F. and {Le Fustec}, Y. and {Lebreton}, Y. and {Lebzelter}, T. and {Leccia}, S. and {Leclerc}, N. and {Lecoeur-Taibi}, I. and {Liao}, S. and {Licata}, E. and {Lindstr{\o}m}, E.~P. and {Lister}, T.~A. and {Livanou}, E. and {Lobel}, A. and {Madrero Pardo}, P. and {Managau}, S. and {Mann}, R.~G. and {Marchant}, J.~M. and {Marconi}, M. and {Marcos Santos}, M.~M.~S. and {Marinoni}, S. and {Marocco}, F. and {Marshall}, D.~J. and {Martin Polo}, L. and {Mart{\'\i}n-Fleitas}, J.~M. and {Masip}, A. and {Massari}, D. and {Mastrobuono-Battisti}, A. and {Mazeh}, T. and {McMillan}, P.~J. and {Messina}, S. and {Michalik}, D. and {Millar}, N.~R. and {Mints}, A. and {Molina}, D. and {Molinaro}, R. and {Moln{\'a}r}, L. and {Montegriffo}, P. and {Mor}, R. and {Morbidelli}, R. and {Morel}, T. and {Morris}, D. and {Mulone}, A.~F. and {Munoz}, D. and {Muraveva}, T. and {Murphy}, C.~P. and {Musella}, I. and {Noval}, L. and {Ord{\'e}novic}, C. and {Orr{\`u}}, G. and {Osinde}, J. and {Pagani}, C. and {Pagano}, I. and {Palaversa}, L. and {Palicio}, P.~A. and {Panahi}, A. and {Pawlak}, M. and {Pe{\~n}alosa Esteller}, X. and {Penttil{\"a}}, A. and {Piersimoni}, A.~M. and {Pineau}, F. -X. and {Plachy}, E. and {Plum}, G. and {Poggio}, E. and {Poretti}, E. and {Poujoulet}, E. and {Pr{\v{s}}a}, A. and {Pulone}, L. and {Racero}, E. and {Ragaini}, S. and {Rainer}, M. and {Raiteri}, C.~M. and {Rambaux}, N. and {Ramos}, P. and {Ramos-Lerate}, M. and {Re Fiorentin}, P. and {Regibo}, S. and {Reyl{\'e}}, C. and {Ripepi}, V. and {Riva}, A. and {Rixon}, G. and {Robichon}, N. and {Robin}, C. and {Roelens}, M. and {Rohrbasser}, L. and {Romero-G{\'o}mez}, M. and {Rowell}, N. and {Royer}, F. and {Rybicki}, K.~A. and {Sadowski}, G. and {Sagrist{\`a} Sell{\'e}s}, A. and {Sahlmann}, J. and {Salgado}, J. and {Salguero}, E. and {Samaras}, N. and {Sanchez Gimenez}, V. and {Sanna}, N. and {Santove{\~n}a}, R. and {Sarasso}, M. and {Schultheis}, M. and {Sciacca}, E. and {Segol}, M. and {Segovia}, J.~C. and {S{\'e}gransan}, D. and {Semeux}, D. and {Shahaf}, S. and {Siddiqui}, H.~I. and {Siebert}, A. and {Siltala}, L. and {Slezak}, E. and {Smart}, R.~L. and {Solano}, E. and {Solitro}, F. and {Souami}, D. and {Souchay}, J. and {Spagna}, A. and {Spoto}, F. and {Steele}, I.~A. and {Steidelm{\"u}ller}, H. and {Stephenson}, C.~A. and {S{\"u}veges}, M. and {Szabados}, L. and {Szegedi-Elek}, E. and {Taris}, F. and {Tauran}, G. and {Taylor}, M.~B. and {Teixeira}, R. and {Thuillot}, W. and {Tonello}, N. and {Torra}, F. and {Torra}, J. and {Turon}, C. and {Unger}, N. and {Vaillant}, M. and {van Dillen}, E. and {Vanel}, O. and {Vecchiato}, A. and {Viala}, Y. and {Vicente}, D. and {Voutsinas}, S. and {Weiler}, M. and {Wevers}, T. and {Wyrzykowski}, {\L}. and {Yoldas}, A. and {Yvard}, P. and {Zhao}, H. and {Zorec}, J. and {Zucker}, S. and {Zurbach}, C. and {Zwitter}, T.},
        title = "{Gaia Early Data Release 3. Summary of the contents and survey properties}",
      journal = {A\&A},
         year = 2021,
        month = may,
       volume = {649},
          doi = {10.1051/0004-6361/202039657},
       adsurl = {https://ui.adsabs.harvard.edu/abs/2021A&A...649A...1G}
}

@ARTICLE{gaia23,
       author = {{Gaia Collaboration} and {Vallenari}, A. and {Brown}, A.~G.~A. and {Prusti}, T. and {de Bruijne}, J.~H.~J. and {Arenou}, F. and {Babusiaux}, C. and {Biermann}, M. and {Creevey}, O.~L. and {Ducourant}, C. and {Evans}, D.~W. and {Eyer}, L. and {Guerra}, R. and {Hutton}, A. and {Jordi}, C. and {Klioner}, S.~A. and {Lammers}, U.~L. and {Lindegren}, L. and {Luri}, X. and {Mignard}, F. and {Panem}, C. and {Pourbaix}, D. and {Randich}, S. and {Sartoretti}, P. and {Soubiran}, C. and {Tanga}, P. and {Walton}, N.~A. and {Bailer-Jones}, C.~A.~L. and {Bastian}, U. and {Drimmel}, R. and {Jansen}, F. and {Katz}, D. and {Lattanzi}, M.~G. and {van Leeuwen}, F. and {Bakker}, J. and {Cacciari}, C. and {Casta{\~n}eda}, J. and {De Angeli}, F. and {Fabricius}, C. and {Fouesneau}, M. and {Fr{\'e}mat}, Y. and {Galluccio}, L. and {Guerrier}, A. and {Heiter}, U. and {Masana}, E. and {Messineo}, R. and {Mowlavi}, N. and {Nicolas}, C. and {Nienartowicz}, K. and {Pailler}, F. and {Panuzzo}, P. and {Riclet}, F. and {Roux}, W. and {Seabroke}, G.~M. and {Sordo}, R. and {Th{\'e}venin}, F. and {Gracia-Abril}, G. and {Portell}, J. and {Teyssier}, D. and {Altmann}, M. and {Andrae}, R. and {Audard}, M. and {Bellas-Velidis}, I. and {Benson}, K. and {Berthier}, J. and {Blomme}, R. and {Burgess}, P.~W. and {Busonero}, D. and {Busso}, G. and {C{\'a}novas}, H. and {Carry}, B. and {Cellino}, A. and {Cheek}, N. and {Clementini}, G. and {Damerdji}, Y. and {Davidson}, M. and {de Teodoro}, P. and {Nu{\~n}ez Campos}, M. and {Delchambre}, L. and {Dell'Oro}, A. and {Esquej}, P. and {Fern{\'a}ndez-Hern{\'a}ndez}, J. and {Fraile}, E. and {Garabato}, D. and {Garc{\'\i}a-Lario}, P. and {Gosset}, E. and {Haigron}, R. and {Halbwachs}, J.-L. and {Hambly}, N.~C. and {Harrison}, D.~L. and {Hern{\'a}ndez}, J. and {Hestroffer}, D. and {Hodgkin}, S.~T. and {Holl}, B. and {Jan{\ss}en}, K. and {Jevardat de Fombelle}, G. and {Jordan}, S. and {Krone-Martins}, A. and {Lanzafame}, A.~C. and {L{\"o}ffler}, W. and {Marchal}, O. and {Marrese}, P.~M. and {Moitinho}, A. and {Muinonen}, K. and {Osborne}, P. and {Pancino}, E. and {Pauwels}, T. and {Recio-Blanco}, A. and {Reyl{\'e}}, C. and {Riello}, M. and {Rimoldini}, L. and {Roegiers}, T. and {Rybizki}, J. and {Sarro}, L.~M. and {Siopis}, C. and {Smith}, M. and {Sozzetti}, A. and {Utrilla}, E. and {van Leeuwen}, M. and {Abbas}, U. and {{\'A}brah{\'a}m}, P. and {Abreu Aramburu}, A. and {Aerts}, C. and {Aguado}, J.~J. and {Ajaj}, M. and {Aldea-Montero}, F. and {Altavilla}, G. and {{\'A}lvarez}, M.~A. and {Alves}, J. and {Anders}, F. and {Anderson}, R.~I. and {Anglada Varela}, E. and {Antoja}, T. and {Baines}, D. and {Baker}, S.~G. and {Balaguer-N{\'u}{\~n}ez}, L. and {Balbinot}, E. and {Balog}, Z. and {Barache}, C. and {Barbato}, D. and {Barros}, M. and {Barstow}, M.~A. and {Bartolom{\'e}}, S. and {Bassilana}, J.-L. and {Bauchet}, N. and {Becciani}, U. and {Bellazzini}, M. and {Berihuete}, A. and {Bernet}, M. and {Bertone}, S. and {Bianchi}, L. and {Binnenfeld}, A. and {Blanco-Cuaresma}, S. and {Blazere}, A. and {Boch}, T. and {Bombrun}, A. and {Bossini}, D. and {Bouquillon}, S. and {Bragaglia}, A. and {Bramante}, L. and {Breedt}, E. and {Bressan}, A. and {Brouillet}, N. and {Brugaletta}, E. and {Bucciarelli}, B. and {Burlacu}, A. and {Butkevich}, A.~G. and {Buzzi}, R. and {Caffau}, E. and {Cancelliere}, R. and {Cantat-Gaudin}, T. and {Carballo}, R. and {Carlucci}, T. and {Carnerero}, M.~I. and {Carrasco}, J.~M. and {Casamiquela}, L. and {Castellani}, M. and {Castro-Ginard}, A. and {Chaoul}, L. and {Charlot}, P. and {Chemin}, L. and {Chiaramida}, V. and {Chiavassa}, A. and {Chornay}, N. and {Comoretto}, G. and {Contursi}, G. and {Cooper}, W.~J. and {Cornez}, T. and {Cowell}, S. and {Crifo}, F. and {Cropper}, M. and {Crosta}, M. and {Crowley}, C. and {Dafonte}, C. and {Dapergolas}, A. and {David}, M. and {David}, P. and {de Laverny}, P. and {De Luise}, F. and {De March}, R.},
        title = "{Gaia Data Release 3. Summary of the content and survey properties}",
      journal = {\aap},
         year = 2023,
        month = jun,
       volume = {674},
          eid = {A1},
        pages = {A1},
          doi = {10.1051/0004-6361/202243940},
archivePrefix = {arXiv},
       eprint = {2208.00211},
 primaryClass = {astro-ph.GA},
       adsurl = {https://ui.adsabs.harvard.edu/abs/2023A&A...674A...1G}
}

@INPROCEEDINGS{greisen03,
       author = {{Greisen}, E.~W.},
        title = "{AIPS, the VLA, and the VLBA}",
    booktitle = {Information Handling in Astronomy - Historical Vistas},
         year = 2003,
       editor = {{Heck}, Andr{\'e}},
       series = {Astrophysics and Space Science Library},
       volume = {285},
        month = mar,
        pages = {109},
          doi = {10.1007/0-306-48080-8_7},
       adsurl = {https://ui.adsabs.harvard.edu/abs/2003ASSL..285..109G}
}

@ARTICLE{gunther22,
       author = {{G{\"u}nther}, Maximilian N. and {Berardo}, David A. and {Ducrot}, Elsa and {Murray}, Catriona A. and {Stassun}, Keivan G. and {Olah}, Katalin and {Bouma}, L.~G. and {Rappaport}, Saul and {Winn}, Joshua N. and {Feinstein}, Adina D. and {Matthews}, Elisabeth C. and {Sebastian}, Daniel and {Rackham}, Benjamin V. and {Seli}, B{\'a}lint and {Triaud}, Amaury H.~M.~J. and {Gillen}, Edward and {Levine}, Alan M. and {Demory}, Brice-Olivier and {Gillon}, Micha{\"e}l and {Queloz}, Didier and {Ricker}, George R. and {Vanderspek}, Roland K. and {Seager}, Sara and {Latham}, David W. and {Jenkins}, Jon M. and {Brasseur}, C.~E. and {Col{\'o}n}, Knicole D. and {Daylan}, Tansu and {Delrez}, Laetitia and {Fausnaugh}, Michael and {Garcia}, Lionel J. and {Jayaraman}, Rahul and {Jehin}, Emmanuel and {Kristiansen}, Martti H. and {Kruijssen}, J.~M. Diederik and {Pedersen}, Peter Pihlmann and {Pozuelos}, Francisco J. and {Rodriguez}, Joseph E. and {Wohler}, Bill and {Zhan}, Zhuchang},
        title = "{Complex Modulation of Rapidly Rotating Young M Dwarfs: Adding Pieces to the Puzzle}",
      journal = {\aj},
         year = 2022,
        month = apr,
       volume = {163},
       number = {4},
          eid = {144},
        pages = {144},
          doi = {10.3847/1538-3881/ac503c},
archivePrefix = {arXiv},
       eprint = {2008.11681},
 primaryClass = {astro-ph.SR},
       adsurl = {https://ui.adsabs.harvard.edu/abs/2022AJ....163..144G}
}

@ARTICLE{hunter07,
       author = {{Hunter}, John D.},
        title = "{Matplotlib: A 2D Graphics Environment}",
      journal = {Computing in Science and Engineering},
         year = 2007,
        month = jan,
       volume = {9},
       number = {3},
        pages = {90-95},
          doi = {10.1109/MCSE.2007.55},
       adsurl = {https://ui.adsabs.harvard.edu/abs/2007CSE.....9...90H}
}

@ARTICLE{kao16,
       author = {{Kao}, Melodie M. and {Hallinan}, Gregg and {Pineda}, J. Sebastian and {Escala}, Ivanna and {Burgasser}, Adam and {Bourke}, Stephen and {Stevenson}, David},
        title = "{Auroral Radio Emission from Late L and T Dwarfs: A New Constraint on Dynamo Theory in the Substellar Regime}",
      journal = {\apj},
         year = 2016,
        month = feb,
       volume = {818},
       number = {1},
          eid = {24},
        pages = {24},
          doi = {10.3847/0004-637X/818/1/24},
archivePrefix = {arXiv},
       eprint = {1511.03661},
 primaryClass = {astro-ph.SR},
       adsurl = {https://ui.adsabs.harvard.edu/abs/2016ApJ...818...24K}
}

@ARTICLE{kao18,
       author = {{Kao}, Melodie M. and {Hallinan}, Gregg and {Pineda}, J. Sebastian and {Stevenson}, David and {Burgasser}, Adam},
        title = "{The Strongest Magnetic Fields on the Coolest Brown Dwarfs}",
      journal = {\apjs},
         year = 2018,
        month = aug,
       volume = {237},
       number = {2},
          eid = {25},
        pages = {25},
          doi = {10.3847/1538-4365/aac2d5},
archivePrefix = {arXiv},
       eprint = {1808.02485},
 primaryClass = {astro-ph.SR},
       adsurl = {https://ui.adsabs.harvard.edu/abs/2018ApJS..237...25K}
}

@ARTICLE{kao23,
       author = {{Kao}, Melodie M. and {Mioduszewski}, Amy J. and {Villadsen}, Jackie and {Shkolnik}, Evgenya L.},
        title = "{Resolved imaging confirms a radiation belt around an ultracool dwarf}",
      journal = {\nat},
         year = 2023,
        month = jul,
       volume = {619},
       number = {7969},
        pages = {272-275},
          doi = {10.1038/s41586-023-06138-w},
archivePrefix = {arXiv},
       eprint = {2302.12841},
 primaryClass = {astro-ph.EP},
       adsurl = {https://ui.adsabs.harvard.edu/abs/2023Natur.619..272K}
}

@ARTICLE{kaur24,
       author = {{Kaur}, Simranpreet and {Vigan{\`o}}, Daniele and {B{\'e}jar}, V{\'\i}ctor J.~S. and {S{\'a}nchez Monge}, {\'A}lvaro and {Morata}, {\`O}scar and {Kansabanik}, Devojyoti and {Girart}, Josep Miquel and {Morales}, Juan Carlos and {Anglada-Escud{\'e}}, Guillem and {Murgas}, Felipe and {Shan}, Yutong and {Ilin}, Ekaterina and {P{\'e}rez-Torres}, Miguel and {Zapatero Osorio}, Mar{\'\i}a Rosa and {Amado}, Pedro J. and {Caballero}, Jos{\'e} A. and {Del Sordo}, Fabio and {Palle}, Enric and {Quirrenbach}, Andreas and {Reiners}, Ansgar and {Ribas}, Ignasi},
        title = "{Hints of auroral and magnetospheric polarized radio emission from the scallop-shell star 2MASS J05082729{\textendash}2101444}",
      journal = {\aap},
         year = 2024,
        month = nov,
       volume = {691},
          eid = {L17},
        pages = {L17},
          doi = {10.1051/0004-6361/202452037},
archivePrefix = {arXiv},
       eprint = {2410.22449},
 primaryClass = {astro-ph.SR},
       adsurl = {https://ui.adsabs.harvard.edu/abs/2024A&A...691L..17K}
}

@ARTICLE{kaur25,
       author = {{Kaur}, Simranpreet and {Vigan{\`o}}, Daniele and {Villadsen}, Jackie and {Miquel Girart}, Josep and {B{\'e}jar}, V{\'\i}ctor J.~S. and {Shan}, Yutong and {Bouma}, Luke and {Ilin}, Ekaterina and {Morata}, {\`O}scar and {P{\'e}rez-Torres}, Miguel and {Bonnassieux}, Etienne and {Gherson}, Jorge R.},
        title = "{Polarized, variable radio emission from the scallop-shell binary system DG CVn}",
      journal = {\aap},
         year = 2025,
        month = sep,
       volume = {701},
          eid = {A69},
        pages = {A69},
          doi = {10.1051/0004-6361/202555222},
archivePrefix = {arXiv},
       eprint = {2507.09366},
 primaryClass = {astro-ph.SR},
       adsurl = {https://ui.adsabs.harvard.edu/abs/2025A&A...701A..69K}
}

@ARTICLE{kaur26,
       author = {{Kaur}, Simranpreet and {Vigan{\`o}}, Daniele and {Girart}, Josep Miquel and {B{\'e}jar}, V{\'\i}ctor J.~S. and {et al.}},
      journal = {in prep.},
year = 2026
}

@ARTICLE{kraus12,
       author = {{Kraus}, Adam L. and {Ireland}, Michael J. and {Hillenbrand}, Lynne A. and {Martinache}, Frantz},
        title = "{The Role of Multiplicity in Disk Evolution and Planet Formation}",
      journal = {\apj},
         year = 2012,
        month = jan,
       volume = {745},
       number = {1},
          eid = {19},
        pages = {19},
          doi = {10.1088/0004-637X/745/1/19},
archivePrefix = {arXiv},
       eprint = {1109.4141},
 primaryClass = {astro-ph.EP},
       adsurl = {https://ui.adsabs.harvard.edu/abs/2012ApJ...745...19K}
}

@ARTICLE{launhardt22,
       author = {{Launhardt}, Ralf and {Loinard}, Laurent and {Dzib}, Sergio A. and {Forbrich}, Jan and {Bower}, Geoffrey C. and {Henning}, Thomas K. and {Mioduszewski}, Amy J. and {Reffert}, Sabine},
        title = "{Nonthermal Radio Continuum Emission from Young Nearby Stars}",
      journal = {\apj},
         year = 2022,
        month = may,
       volume = {931},
       number = {1},
          eid = {43},
        pages = {43},
          doi = {10.3847/1538-4357/ac5b09},
archivePrefix = {arXiv},
       eprint = {2203.03418},
 primaryClass = {astro-ph.SR},
       adsurl = {https://ui.adsabs.harvard.edu/abs/2022ApJ...931...43L}
}

@ARTICLE{leto21,
       author = {{Leto}, P. and {Trigilio}, C. and {Krti{\v{c}}ka}, J. and {Fossati}, L. and {Ignace}, R. and {Shultz}, M.~E. and {Buemi}, C.~S. and {Cerrigone}, L. and {Umana}, G. and {Ingallinera}, A. and {Bordiu}, C. and {Pillitteri}, I. and {Bufano}, F. and {Oskinova}, L.~M. and {Agliozzo}, C. and {Cavallaro}, F. and {Riggi}, S. and {Loru}, S. and {Todt}, H. and {Giarrusso}, M. and {Phillips}, N.~M. and {Robrade}, J. and {Leone}, F.},
        title = "{A scaling relationship for non-thermal radio emission from ordered magnetospheres: from the top of the main sequence to planets}",
      journal = {\mnras},
         year = 2021,
        month = oct,
       volume = {507},
       number = {2},
        pages = {1979-1998},
          doi = {10.1093/mnras/stab2168},
archivePrefix = {arXiv},
       eprint = {2107.11995},
 primaryClass = {astro-ph.SR},
       adsurl = {https://ui.adsabs.harvard.edu/abs/2021MNRAS.507.1979L}
}

@ARTICLE{loinard07,
       author = {{Loinard}, Laurent and {Rodr{\'\i}guez}, Luis F. and {D'Alessio}, Paola and {Rodr{\'\i}guez}, M{\'o}nica I. and {Gonz{\'a}lez}, Ricardo F.},
        title = "{On the Nature of the Extended Radio Emission Surrounding T Tauri South}",
      journal = {\apj},
         year = 2007,
        month = mar,
       volume = {657},
       number = {2},
        pages = {916-924},
          doi = {10.1086/510994},
       adsurl = {https://ui.adsabs.harvard.edu/abs/2007ApJ...657..916L}
}

@ARTICLE{lynch16,
       author = {{Lynch}, C. and {Murphy}, T. and {Ravi}, V. and {Hobbs}, G. and {Lo}, K. and {Ward}, C.},
        title = "{Radio detections of southern ultracool dwarfs}",
      journal = {\mnras},
         year = 2016,
        month = apr,
       volume = {457},
       number = {2},
        pages = {1224-1232},
          doi = {10.1093/mnras/stw050},
archivePrefix = {arXiv},
       eprint = {1601.01749},
 primaryClass = {astro-ph.SR},
       adsurl = {https://ui.adsabs.harvard.edu/abs/2016MNRAS.457.1224L}
}

@ARTICLE{melrose82,
       author = {{Melrose}, D.~B. and {Dulk}, G.~A.},
        title = "{Electron-cyclotron masers as the source of certain solar and stellar radio bursts.}",
      journal = {\apj},
         year = 1982,
        month = aug,
       volume = {259},
        pages = {844-858},
          doi = {10.1086/160219},
       adsurl = {https://ui.adsabs.harvard.edu/abs/1982ApJ...259..844M}
}

@ARTICLE{miret20,
       author = {{Miret-Roig}, N. and {Galli}, P.~A.~B. and {Brandner}, W. and {Bouy}, H. and {Barrado}, D. and {Olivares}, J. and {Antoja}, T. and {Romero-G{\'o}mez}, M. and {Figueras}, F. and {Lillo-Box}, J.},
        title = "{Dynamical traceback age of the {\ensuremath{\beta}} Pictoris moving group}",
      journal = {\aap},
         year = 2020,
        month = oct,
       volume = {642},
          eid = {A179},
        pages = {A179},
          doi = {10.1051/0004-6361/202038765},
archivePrefix = {arXiv},
       eprint = {2007.10997},
 primaryClass = {astro-ph.GA},
       adsurl = {https://ui.adsabs.harvard.edu/abs/2020A&A...642A.179M}
}

@software{newville20,
       author = {{Newville}, Matt and {Otten}, Renee and {Nelson}, Andrew and {Ingargiola}, Antonino and {Stensitzki}, Till and {Allan}, Dan and {Fox}, Austin and {Carter}, Faustin and {Micha{\l}} and {Pustakhod}, Dima and {Ram}, Yoav and {Glenn} and {Deil}, Christoph and {Stuermer} and {Beelen}, Alexandre and {Frost}, Oliver and {Zobrist}, Nicholas and {Mark} and {Pasquevich}, Gustavo and {Hansen}, Allan L.~R. and {Spillane}, Tim and {Caldwell}, Shane and {Polloreno}, Anthony and {Andrewhannum} and {Fraine}, Jonathan and {Deep-42-Thought} and {Maier}, Benjamin F. and {Gamari}, Ben and {Persaud}, Arun and {Almarza}, Anthony},
        title = "{lmfit/lmfit-py 1.0.1}",
         year = 2020,
        month = may,
          eid = {10.5281/zenodo.3814709},
          doi = {10.5281/zenodo.3814709},
      version = {1.0.1},
    publisher = {Zenodo},
       adsurl = {https://ui.adsabs.harvard.edu/abs/2020zndo...3814709N}
}

@ARTICLE{ortizleon17,
       author = {{Ortiz-Le{\'o}n}, Gisela N. and {Loinard}, Laurent and {Kounkel}, Marina A. and {Dzib}, Sergio A. and {Mioduszewski}, Amy J. and {Rodr{\'\i}guez}, Luis F. and {Torres}, Rosa M. and {Gonz{\'a}lez-L{\'o}pezlira}, Rosa A. and {Pech}, Gerardo and {Rivera}, Juana L. and {Hartmann}, Lee and {Boden}, Andrew F. and {Evans}, II, Neal J. and {Brice{\~n}o}, Cesar and {Tobin}, John J. and {Galli}, Phillip A.~B. and {Gudehus}, Donald},
        title = "{The Gould{\textquoteright}s Belt Distances Survey (GOBELINS). I. Trigonometric Parallax Distances and Depth of the Ophiuchus Complex}",
      journal = {\apj},
         year = 2017,
        month = jan,
       volume = {834},
       number = {2},
          eid = {141},
        pages = {141},
          doi = {10.3847/1538-4357/834/2/141},
archivePrefix = {arXiv},
       eprint = {1611.06466},
 primaryClass = {astro-ph.SR},
       adsurl = {https://ui.adsabs.harvard.edu/abs/2017ApJ...834..141O}
}

@ARTICLE{parsamyan95,
       author = {{Parsamyan}, E.~S.},
        title = "{Determination of the age of stellar aggregates and flare stars of the galactic field}",
      journal = {Astrophysics},
         year = 1995,
        month = jul,
       volume = {38},
       number = {3},
        pages = {206-212},
          doi = {10.1007/BF02045328},
       adsurl = {https://ui.adsabs.harvard.edu/abs/1995Ap.....38..206P}
}

@ARTICLE{pineda17,
       author = {{Pineda}, J. Sebastian and {Hallinan}, Gregg and {Kao}, Melodie M.},
        title = "{A Panchromatic View of Brown Dwarf Aurorae}",
      journal = {\apj},
         year = 2017,
        month = sep,
       volume = {846},
       number = {1},
          eid = {75},
        pages = {75},
          doi = {10.3847/1538-4357/aa8596},
archivePrefix = {arXiv},
       eprint = {1708.02942},
 primaryClass = {astro-ph.SR},
       adsurl = {https://ui.adsabs.harvard.edu/abs/2017ApJ...846...75P}
}

@ARTICLE{pritchard21,
       author = {{Pritchard}, Joshua and {Murphy}, Tara and {Zic}, Andrew and {Lynch}, Christene and {Heald}, George and {Kaplan}, David L. and {Anderson}, Craig and {Banfield}, Julie and {Hale}, Catherine and {Hotan}, Aidan and {Lenc}, Emil and {Leung}, James K. and {McConnell}, David and {Moss}, Vanessa A. and {Raja}, Wasim and {Stewart}, Adam J. and {Whiting}, Matthew},
        title = "{A circular polarization survey for radio stars with the Australian SKA Pathfinder}",
      journal = {\mnras},
         year = 2021,
        month = apr,
       volume = {502},
       number = {4},
        pages = {5438-5454},
          doi = {10.1093/mnras/stab299},
archivePrefix = {arXiv},
       eprint = {2102.01801},
 primaryClass = {astro-ph.SR},
       adsurl = {https://ui.adsabs.harvard.edu/abs/2021MNRAS.502.5438P}
}

@ARTICLE{pritchard24,
       author = {{Pritchard}, Joshua and {Murphy}, Tara and {Heald}, George and {Wheatland}, Michael S. and {Kaplan}, David L. and {Lenc}, Emil and {O'Brien}, Andrew and {Wang}, Ziteng},
        title = "{Multi-epoch sampling of the radio star population with the Australian SKA Pathfinder}",
      journal = {\mnras},
         year = 2024,
        month = apr,
       volume = {529},
       number = {2},
        pages = {1258-1270},
          doi = {10.1093/mnras/stae127},
archivePrefix = {arXiv},
       eprint = {2312.11031},
 primaryClass = {astro-ph.SR},
       adsurl = {https://ui.adsabs.harvard.edu/abs/2024MNRAS.529.1258P}
}

@INPROCEEDINGS{quirrenbach14,
       author = {{Quirrenbach}, A. and {Amado}, P.~J. and {Caballero}, J.~A. and {Mundt}, R. and {Reiners}, A. and {Ribas}, I. and {Seifert}, W. and {Abril}, M. and {Aceituno}, J. and {Alonso-Floriano}, F.~J. and et al.},
        title = "{CARMENES instrument overview}",
    booktitle = {Ground-based and Airborne Instrumentation for Astronomy V},
         year = 2014,
       editor = {{Ramsay}, Suzanne K. and {McLean}, Ian S. and {Takami}, Hideki},
       series = {Society of Photo-Optical Instrumentation Engineers (SPIE) Conference Series},
       volume = {9147},
        month = jul,
          eid = {91471F},
        pages = {91471F},
          doi = {10.1117/12.2056453},
       adsurl = {https://ui.adsabs.harvard.edu/abs/2014SPIE.9147E..1FQ}
}

@INPROCEEDINGS{quirrenbach16,
       author = {{Quirrenbach}, A. and {Amado}, P.~J. and {Caballero}, J.~A. and {Mundt}, R. and {Reiners}, A. and {Ribas}, I. and {Seifert}, W. and {Abril}, M. and {Aceituno}, J. and {Alonso-Floriano}, F.~J. and {Anwand-Heerwart}, H. and {Azzaro}, M. and {Bauer}, F. and {Barrado}, D. and {Becerril}, S. and {Bejar}, V.~J.~S. and {Benitez}, D. and {Berdinas}, Z.~M. and {Brinkm{\"o}ller}, M. and {Cardenas}, M.~C. and {Casal}, E. and {Claret}, A. and {Colom{\'e}}, J. and {Cortes-Contreras}, M. and {Czesla}, S. and {Doellinger}, M. and {Dreizler}, S. and {Feiz}, C. and {Fernandez}, M. and {Ferro}, I.~M. and {Fuhrmeister}, B. and {Galadi}, D. and {Gallardo}, I. and {G{\'a}lvez-Ortiz}, M.~C. and {Garcia-Piquer}, A. and {Garrido}, R. and {Gesa}, L. and {G{\'o}mez Galera}, V. and {Gonz{\'a}lez Hern{\'a}ndez}, J.~I. and {Gonzalez Peinado}, R. and {Gr{\"o}zinger}, U. and {Gu{\`a}rdia}, J. and {Guenther}, E.~W. and {de Guindos}, E. and {Hagen}, H. -J. and {Hatzes}, A.~P. and {Hauschildt}, P.~H. and {Helmling}, J. and {Henning}, T. and {Hermann}, D. and {Hern{\'a}ndez Arabi}, R. and {Hern{\'a}ndez Casta{\~n}o}, L. and {Hern{\'a}ndez Hernando}, F. and {Herrero}, E. and {Huber}, A. and {Huber}, K.~F. and {Huke}, P. and {Jeffers}, S.~V. and {de Juan}, E. and {Kaminski}, A. and {Kehr}, M. and {Kim}, M. and {Klein}, R. and {Kl{\"u}ter}, J. and {K{\"u}rster}, M. and {Lafarga}, M. and {Lara}, L.~M. and {Lamert}, A. and {Laun}, W. and {Launhardt}, R. and {Lemke}, U. and {Lenzen}, R. and {Llamas}, M. and {Lopez del Fresno}, M. and {L{\'o}pez-Puertas}, M. and {L{\'o}pez-Santiago}, J. and {Lopez Salas}, J.~F. and {Magan Madinabeitia}, H. and {Mall}, U. and {Mandel}, H. and {Mancini}, L. and {Marin Molina}, J.~A. and {Maroto Fern{\'a}ndez}, D. and {Mart{\'\i}n}, E.~L. and {Mart{\'\i}n-Ruiz}, S. and {Marvin}, C. and {Mathar}, R.~J. and {Mirabet}, E. and {Montes}, D. and {Morales}, J.~C. and {Morales Mu{\~n}oz}, R. and {Nagel}, E. and {Naranjo}, V. and {Nowak}, G. and {Palle}, E. and {Panduro}, J. and {Passegger}, V.~M. and {Pavlov}, A. and {Pedraz}, S. and {Perez}, E. and {P{\'e}rez-Medialdea}, D. and {Perger}, M. and {Pluto}, M. and {Ram{\'o}n}, A. and {Rebolo}, R. and {Redondo}, P. and {Reffert}, S. and {Reinhart}, S. and {Rhode}, P. and {Rix}, H. -W. and {Rodler}, F. and {Rodr{\'\i}guez}, E. and {Rodr{\'\i}guez L{\'o}pez}, C. and {Rohloff}, R.~R. and {Rosich}, A. and {Sanchez Carrasco}, M.~A. and {Sanz-Forcada}, J. and {Sarkis}, P. and {Sarmiento}, L.~F. and {Sch{\"a}fer}, S. and {Schiller}, J. and {Schmidt}, C. and {Schmitt}, J.~H.~M.~M. and {Sch{\"o}fer}, P. and {Schweitzer}, A. and {Shulyak}, D. and {Solano}, E. and {Stahl}, O. and {Storz}, C. and {Tabernero}, H.~M. and {Tala}, M. and {Tal-Or}, L. and {Ulbrich}, R. -G. and {Veredas}, G. and {Vico Linares}, J.~I. and {Vilardell}, F. and {Wagner}, K. and {Winkler}, J. and {Zapatero Osorio}, M. -R. and {Zechmeister}, M. and {Ammler-von Eiff}, M. and {Anglada-Escud{\'e}}, G. and {del Burgo}, C. and {Garcia-Vargas}, M.~L. and {Klutsch}, A. and {Lizon}, J. -L. and {Lopez-Morales}, M. and {Ofir}, A. and {P{\'e}rez-Calpena}, A. and {Perryman}, M.~A.~C. and {S{\'a}nchez-Blanco}, E. and {Strachan}, J.~B.~P. and {St{\"u}rmer}, J. and {Su{\'a}rez}, J.~C. and {Trifonov}, T. and {Tulloch}, S.~M. and {Xu}, W.},
        title = "{CARMENES: an overview six months after first light}",
    booktitle = {Ground-based and Airborne Instrumentation for Astronomy VI},
         year = 2016,
       editor = {{Evans}, Christopher J. and {Simard}, Luc and {Takami}, Hideki},
       series = {Society of Photo-Optical Instrumentation Engineers (SPIE) Conference Series},
       volume = {9908},
        month = aug,
          eid = {990812},
        pages = {990812},
          doi = {10.1117/12.2231880},
       adsurl = {https://ui.adsabs.harvard.edu/abs/2016SPIE.9908E..12Q}
}

@INPROCEEDINGS{quirrenbach18,
       author = {{Quirrenbach}, A. and {Amado}, P.~J. and {Ribas}, I. and {Reiners}, A. and {Caballero}, J.~A. and {Seifert}, W. and {Aceituno}, J. and {Azzaro}, M. and {Baroch}, D. and {Barrado}, D. and {Bauer}, F. and {Becerril}, S. and {B{\`e}jar}, V.~J.~S. and {Ben{\'\i}tez}, D. and {Brinkm{\"o}ller}, M. and {Cardona Guill{\'e}n}, C. and {Cifuentes}, C. and {Colom{\'e}}, J. and {Cort{\'e}s-Contreras}, M. and {Czesla}, S. and {Dreizler}, S. and {Fr{\"o}lich}, K. and {Fuhrmeister}, B. and {Galad{\'\i}-Enr{\'\i}quez}, D. and {Gonz{\'a}lez Hern{\'a}ndez}, J.~I. and {Gonz{\'a}lez Peinado}, R. and {Guenther}, E.~W. and {de Guindos}, E. and {Hagen}, H.-J. and {Hatzes}, A.~P. and {Hauschildt}, P.~H. and {Helmling}, J. and {Henning}, Th. and {Herbort}, O. and {Hern{\'a}ndez Casta{\~n}o}, L. and {Herrero}, E. and {Hintz}, D. and {Jeffers}, S.~V. and {Johnson}, E.~N. and {de Juan}, E. and {Kaminski}, A. and {Klahr}, H. and {K{\"u}rster}, M. and {Lafarga}, M. and {Sairam}, L. and {Lamp{\'o}n}, M. and {Lara}, L.~M. and {Launhardt}, R. and {L{\'o}pez del Fresno}, M. and {L{\'o}pez-Puertas}, M. and {Luque}, R. and {Mandel}, H. and {Marfil}, E.~G. and {Mart{\'\i}n}, E.~L. and {Mart{\'\i}n-Ruiz}, S. and {Mathar}, R.~J. and {Montes}, D. and {Morales}, J.~C. and {Nagel}, E. and {Nortmann}, L. and {Nowak}, G. and {Pall{\'e}}, E. and {Passegger}, V.-M. and {Pavlov}, A. and {Pedraz}, S. and {P{\'e}rez-Medialdea}, D. and {Perger}, M. and {Rebolo}, R. and {Reffert}, S. and {Rodr{\'\i}guez}, E. and {Rodr{\'\i}guez L{\'o}pez}, C. and {Rosich}, A. and {Sabotta}, S. and {Sadegi}, S. and {Salz}, M. and {S{\'a}nchez-L{\'o}pez}, A. and {Sanz-Forcada}, J. and {Sarkis}, P. and {Sch{\"a}fer}, S. and {Schiller}, J. and {Schmitt}, J.~H.~M.~M. and {Sch{\"o}fer}, P. and {Schweitzer}, A. and {Shulyak}, D. and {Solano}, E. and {Stahl}, O. and {Tala Pinto}, M. and {Trifonov}, T. and {Zapatero Osorio}, M.~R. and {Yan}, F. and {Zechmeister}, M. and {Abell{\'a}n}, F.~J. and {Abril}, M. and {Alonso-Floriano}, F.~J. and {Ammler-von Eiff}, M. and {Anglada-Escud{\'e}}, G. and {Anwand-Heerwart}, H. and {Arroyo-Torres}, B. and {Berdi{\~n}as}, Z.~M. and {Bergondy}, G. and {Bl{\"u}mcke}, M. and {del Burgo}, C. and {Cano}, J. and {Carro}, J. and {C{\'a}rdenas}, M.~C. and {Casal}, E. and {Claret}, A. and {D{\'\i}ez-Alonso}, E. and {Doellinger}, M. and {Dorda}, R. and {Feiz}, C. and {Fern{\'a}ndez}, M. and {Ferro}, I.~M. and {Gaisn{\'e}}, G. and {Gallardo}, I. and {G{\'a}lvez-Ortiz}, M.~C. and {Garc{\'\i}a-Piquer}, A. and {Garc{\'\i}a-Vargas}, M.~L. and {Garrido}, R. and {Gesa}, L. and {G{\'o}mez Galera}, V. and {Gonz{\'a}lez-{\'A}lvarez}, E. and {Gonz{\'a}lez-Cuesta}, L. and {Grohnert}, S. and {Gr{\"o}zinger}, U. and {Gu{\`a}rdia}, J. and {Guijarro}, A. and {Hedrosa}, R.~P. and {Hermann}, D. and {Hermelo}, I. and {Hern{\'a}ndez Arab{\'\i}}, R. and {Hern{\'a}ndez Hernando}, F. and {Hidalgo}, D. and {Holgado}, G. and {Huber}, A. and {Huber}, K. and {Huke}, P. and {Kehr}, M. and {Kim}, M. and {Klein}, R. and {Kl{\"u}ter}, J. and {Klutsch}, A. and {Labarga}, F. and {Labiche}, N. and {Lamert}, A. and {Laun}, W. and {L{\'a}zaro}, F.~J. and {Lemke}, U. and {Lenzen}, R. and {Llamas}, M. and {Lizon}, J.-L. and {Lodieu}, N. and {L{\'o}pez Gonz{\'a}lez}, M.~J. and {L{\'o}pez-Morales}, M. and {L{\'o}pez Salas}, J.~F. and {L{\'o}pez-Santiago}, J. and {Mag{\'a}n Madinabeitia}, H. and {Mall}, U. and {Mancini}, L. and {Mar{\'\i}n Molina}, J.~A. and {Mart{\'\i}nez-Rodr{\'\i}guez}, H. and {Maroto Fern{\'a}ndez}, D. and {Marvin}, C.~J. and {Mirabet}, E. and {Moreno-Raya}, M.~E. and {Moya}, A. and {Mundt}, R. and {Naranjo}, V. and {Panduro}, J. and {Pascual}, J. and {P{\'e}rez-Calpena}, A. and {Perryman}, M.~A.~C. and {Pluto}, M. and {Ram{\'o}n}, A. and {Redondo}, P. and {Reinhart}, S. and {Rhode}, P. and {Rix}, H.-W. and {Rodler}, F. and {Rohloff}, R.-R. and {S{\'a}nchez-Blanco}, E. and {S{\'a}nchez Carrasco}, M.~A. and {Sarmiento}, L.~F. and {Schmidt}, C. and {Storz}, C. and {Strachan}, J.~B.~P. and {St{\"u}rmer}, J. and {Su{\'a}rez}, J.~C. and {Tabernero}, H.~M. and {Tal-Or}, L. and {Tulloch}, S.~M. and {Ulbrich}, R.-G. and {Veredas}, G. and {Vico Linares}, J.~L. and {Vidal-Dasilva}, M. and {Vilardell}, F. and {Wagner}, K. and {Winkler}, J. and {Wolthoff}, V. and {Xu}, W.},
        title = "{CARMENES: high-resolution spectra and precise radial velocities in the red and infrared}",
    booktitle = {Ground-based and Airborne Instrumentation for Astronomy VII},
         year = 2018,
       editor = {{Evans}, Christopher J. and {Simard}, Luc and {Takami}, Hideki},
       series = {Society of Photo-Optical Instrumentation Engineers (SPIE) Conference Series},
       volume = {10702},
        month = jul,
          eid = {107020W},
        pages = {107020W},
          doi = {10.1117/12.2313689},
       adsurl = {https://ui.adsabs.harvard.edu/abs/2018SPIE10702E..0WQ}
}

@ARTICLE{riaz06,
       author = {{Riaz}, Basmah and {Gizis}, John E. and {Harvin}, James},
        title = "{Identification of New M Dwarfs in the Solar Neighborhood}",
      journal = {\aj},
         year = 2006,
        month = aug,
       volume = {132},
       number = {2},
        pages = {866-872},
          doi = {10.1086/505632},
archivePrefix = {arXiv},
       eprint = {astro-ph/0606617},
 primaryClass = {astro-ph},
       adsurl = {https://ui.adsabs.harvard.edu/abs/2006AJ....132..866R}
}

\begin{appendix}
\onecolumn

\section{Corner plot of the combined fit}

\begin{figure}[h!]
\centering
\includegraphics[width=\columnwidth]
{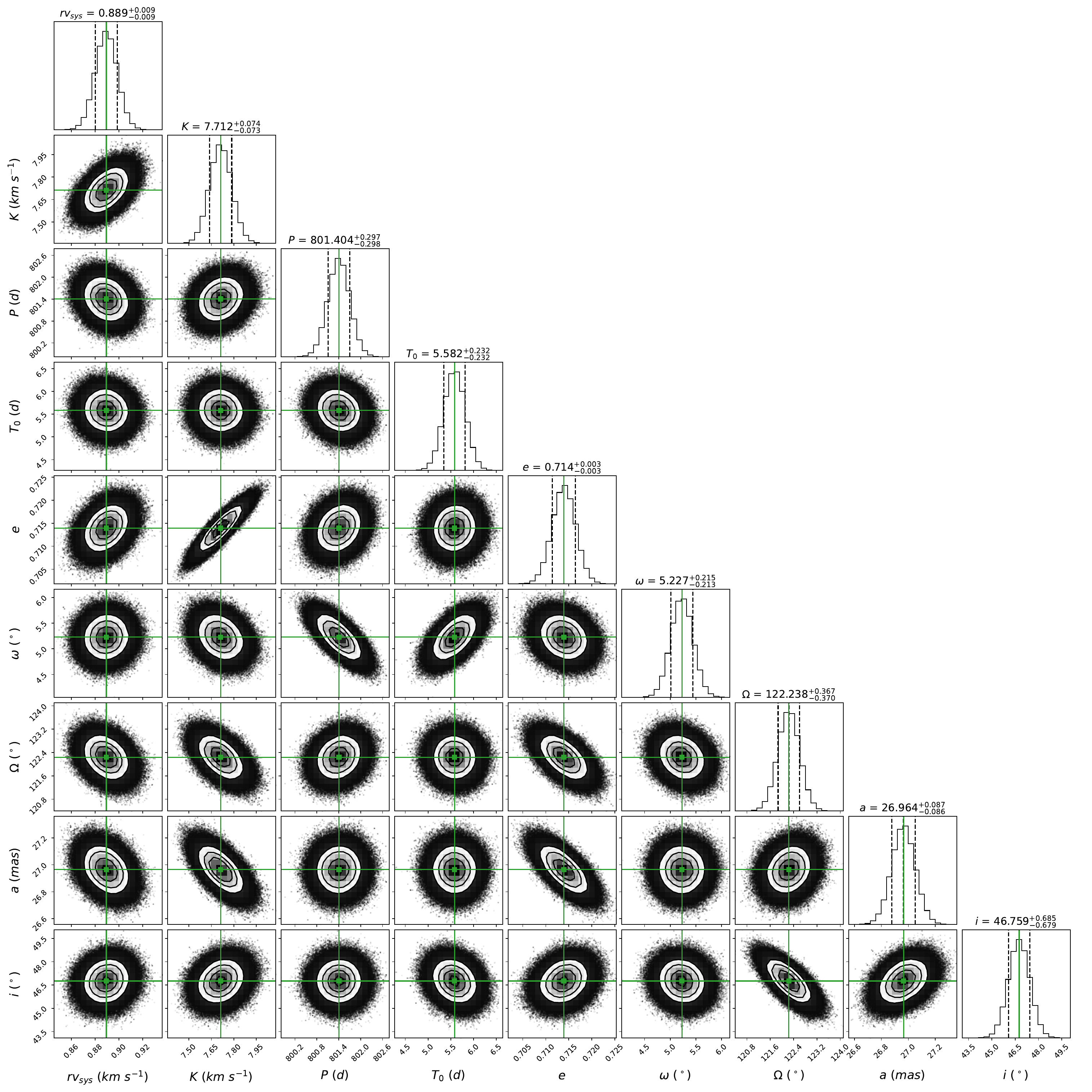}
   \caption{Posterior distributions of the fitted parameters. Combined astrometric fit of the M-dwarf binary system 2M0508--21AB using {\tt lmfit}. This figure shows the correlations between the fitted parameters from the {\tt MCMC} analysis using the corner code. The 2D posterior probability histogram of each fitted parameter is shown on top of each column. The green lines indicate the mean value of each fitted parameter, and the two dotted vertical lines represent the 1$\sigma$ range of the distribution.}
         \label{fig:corner}
   \end{figure}

\section{Individual radio flux density variability and polarization}\label{app:individual_lc}

\begin{figure*}[h!]
\centering
\sidecaption
\includegraphics[width=12cm]
{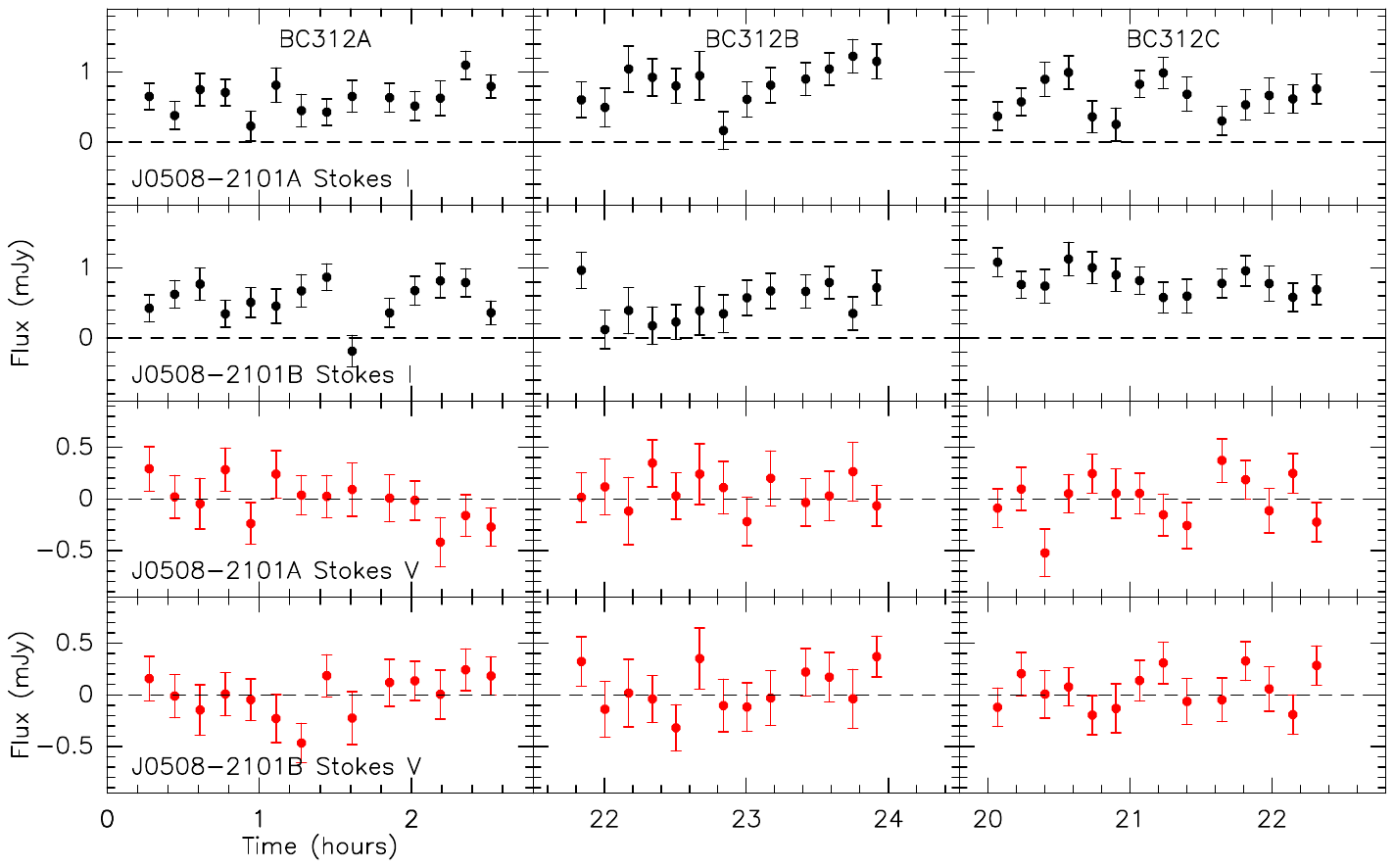} 
\caption{Plots of the flux densities for Stokes I and V of the two components, for the 3 epochs, with time bins of 10 min. These values are estimated in the image plane. In both cases, the upper panels are from the primary star 2M0508--21A, and the lower panels are from the secondary star 2M0508--21B.}
\label{fig:flux_10min}
\end{figure*}

\begin{figure*}[h!]
\centering
\sidecaption
\includegraphics[width=12cm]
{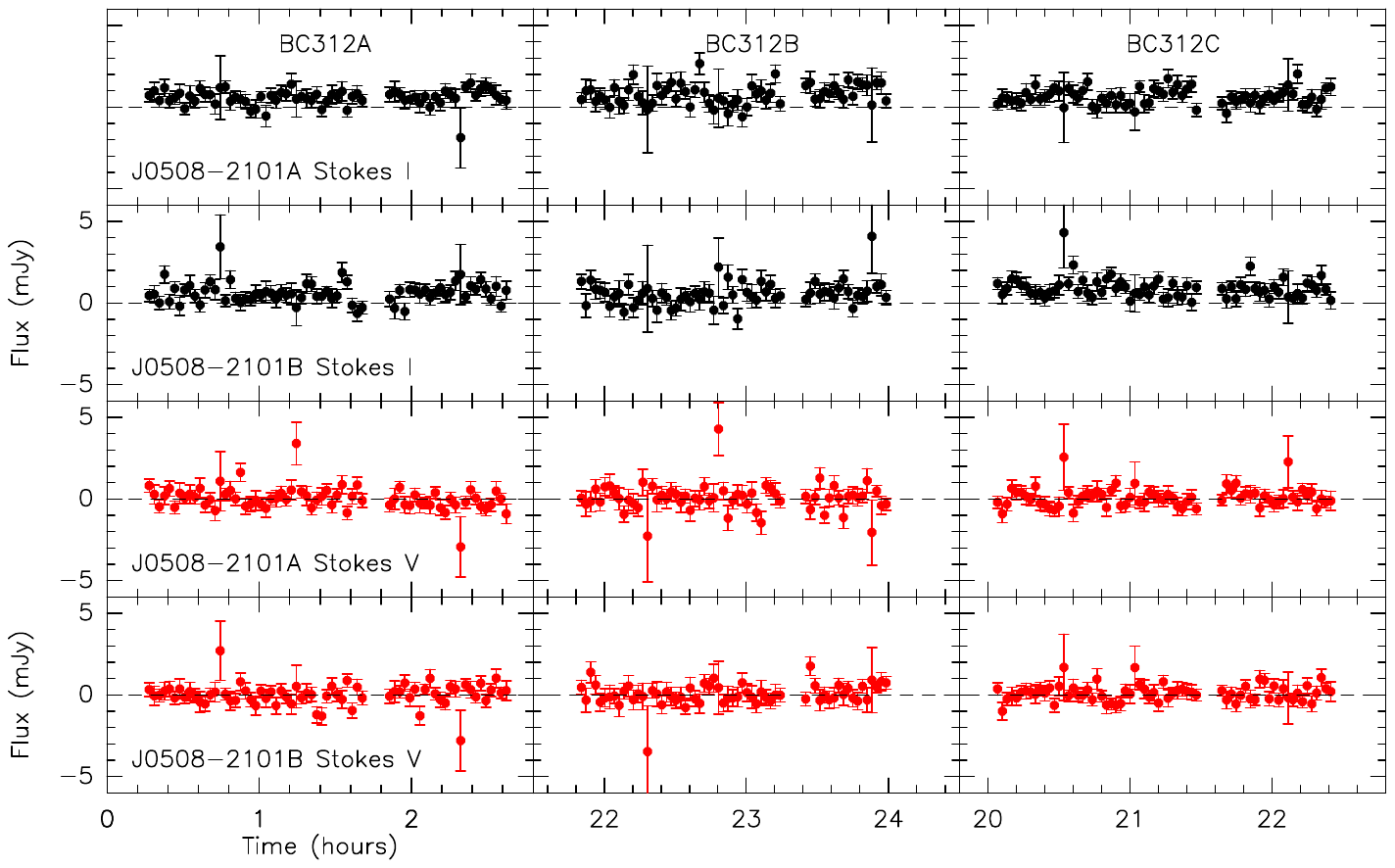}
\caption{Same plot as Fig.~\ref{fig:flux_10min} but for intervals of 2 min.}
\label{fig:flux_2min}
\end{figure*}

To check for possible flares, we have created images with time intervals of 10 and 2 min. The resulting maps have typical rms noise of 0.2 and 0.4~mJy~beam$^{-1}$, respectively (Figs.\,\ref{fig:flux_10min} and Fig.~\ref{fig:flux_2min}). The flux density was estimated using a box with a size about four times larger than the beam size. The 10~min images show some marginal modulation, especially for  2M0508--21A,  during each observation. 
The slight increase in emission in the 2M0508--21A and 2M0508--21B components on the second day (BC312B) is also seen in the visibility shown in Fig.~\ref{fig:vlba_light_curve} (which contains both components). In any case, we can discard significant variability of a factor two or more of the average flux density. 
In the case of the 2 min images, the noise level is higher than the flux density of the two sources, but they can be used to track significant flares. Indeed, there are few points with high flux densities, but these correspond to images with significantly higher rms noise, and therefore the possible flare is doubtful. 
In addition, the poor image fidelity of the 10 and 2 min images prevents us from seeing significant Stokes V emission, if any.

\FloatBarrier 
\clearpage

\end{appendix}
\end{document}